\documentclass[12pt,journal,final,onecolumn]{IEEEtran} 

\usepackage[pdftex]{graphicx}
\usepackage[cmex10]{amsmath}
\usepackage{amssymb}
\usepackage{amsthm}
\usepackage{amsfonts}
\usepackage{mathtools} 
\usepackage{thmtools} 
\usepackage{bm}
\usepackage{cite}
\usepackage[svgnames]{xcolor} 
\usepackage{pstricks,pst-node,pst-plot,pstricks-add}
\usepackage[tight,footnotesize]{subfigure}
\usepackage[binary-units=true,per-mode=symbol]{siunitx}
\usepackage{algorithm}
\usepackage{algpseudocode}
\usepackage{optidef}
\usepackage{hyperref} 

\graphicspath{{figs/}}

\mathtoolsset{showonlyrefs} 

\newcommand{\mc}[1]{\mathcal{#1}}

\newcommand{\msf}[1]{\mathsf{#1}}

\newcommand{\defeq}{\mathrel{\triangleq}}
\newcommand{\Pp}{\mathbb{P}}

\newcommand{\N}{\mathbb{N}}
\newcommand{\Z}{\mathbb{Z}}

\newcommand{\R}{\mathbb{R}}

\newcommand{\C}{\mathbb{C}}

\newcommand{\ind}{1}

\DeclarePairedDelimiter\floor{\lfloor}{\rfloor}
\DeclarePairedDelimiter\abs{\lvert}{\rvert}
\DeclarePairedDelimiter\card{\lvert}{\rvert}
\DeclarePairedDelimiter\norm{\lVert}{\rVert}

\newcommand{\iid}{i.\@i.\@d.\ }

\DeclareMathOperator{\rank}{rank}
\DeclareMathOperator{\spn}{span}
\DeclareMathOperator{\tr}{tr}

\DeclareMathOperator*{\argmax}{arg\,max}

\newtheorem{lemma}{Lemma}

\newtheorem{theorem}[lemma]{Theorem}

\newtheorem{innercustomtheorem}{Theorem}
{\renewcommand\theinnercustomtheorem{#1}\innercustomtheorem}%
{\endinnercustomtheorem}

\theoremstyle{definition}

\declaretheorem[style=definition,qed=$\lozenge$]{example}
\renewcommand\thmcontinues[1]{Continued}

\newtheoremstyle{myremark}%
{\topsep}{\topsep}{\normalfont}{\parindent}{\itshape}{:}{ }{}

\theoremstyle{myremark}
\newtheorem{remark}{Remark}

\begin{document}

\title{Joint Beamforming and Association Design\\ for MIMO Radar} 

\author{Urs Niesen and Jayakrishnan Unnikrishnan%
    \thanks{The authors are with Qualcomm Flarion Technologies, Inc., Bridgewater, NJ 08807, USA. Emails: \{uniesen, junnikri\}@qti.qualcomm.com}%
    \thanks{Part of this paper is presented at the 2018 Asilomar Conference on Signals, Systems, and Computers~\cite{niesen18}.}%
}

\maketitle

\begin{abstract}
    A critical task of a radar receiver is data association, which assigns
    radar target detections to target filter tracks. Motivated by its
    importance, this paper introduces the problem of jointly designing
    multiple-input multiple-output (MIMO) radar transmit beam patterns and the
    corresponding data association schemes. We show that the coupling of the
    beamforming and the association subproblems can be conveniently
    parameterized by what we term an ambiguity graph, which prescribes if two
    targets are to be disambiguated by the beamforming design or by the data
    association scheme.  The choice of ambiguity graph determines which of the
    two subproblems is more difficult and therefore allows to trade performance
    of one versus the other, resulting in a detection-association trade-off.
    This paper shows how to design both the beam pattern and the association
    scheme for a given ambiguity graph. It then discusses how to choose an
    ambiguity graph achieving close to the optimal detection-association
    trade-off.
\end{abstract}

\section{Introduction} 
\label{sec:intro}

\subsection{Motivation and Summary of Results}
\label{sec:intro_motivation}

One of the key advantages of multiple-input multiple-output (MIMO) radar systems
is that they enable adaptive transmit beamforming. This technique adaptively
steers radar beams to better illuminate a target of interest while reducing the
received signal contribution from other targets, jammers, or
clutter~\cite{listoica09}. There has been significant recent interest in
designing these transmit beam patterns, summarized in
Section~\ref{sec:intro_related} below. 

The performance of a transmit beam pattern depends on the corresponding receiver
architecture. A standard radar receiver architecture for tracking targets
consists of two main building blocks: detection of potential targets and track
filtering (see, e.g.,~\cite[Chapter~7.3.4]{richards14},~\cite{barshalom90,
rong03}). The two are linked by an association step, which uses the target
priors to assign each detection to one of the tracks. The association step
ensures that all the detections assigned to the same track are in fact caused by
a single target across time. Correct association is critical, since errors in
this step can lead to tracking failure.

The simplest transmit beamforming methods focus on optimizing coverage within
the field of view. More advanced methods utilize prior information from the
tracker, such as estimates of the targets' azimuth angles. However, to the best
of our knowledge, prior work effectively ignores the association step during the
beamforming design (see Section~\ref{sec:intro_related}).

In this paper, we initiate the joint design of the beam pattern and the data
association scheme. We show that the coupling of the two is captured by what we
term an \emph{ambiguity graph}. The vertices of this graph are the targets. Two
vertices connected by an edge correspond to targets that are ``difficult'' to
disambiguate using the prior distributions during data association. They
therefore should not be simultaneously illuminated by the beam pattern.
Conversely, two vertices not connected by an edge correspond to targets that are
``easy'' to disambiguate during data association. They therefore do not place
any constraints on the beam pattern. The choice of precisely which targets are
``difficult'' or ``easy'' to disambiguate allows to trade the performance of the
beamformer and the association scheme. We refer to this as the
\emph{detection-association trade-off}.

The joint beamforming and association design problem can then be broken up into
three subproblems: First, designing an optimal beam pattern respecting the
constraints of a given ambiguity graph. Second, finding an appropriate association
scheme for a beam pattern with the given ambiguity graph. Third, choosing a
Pareto-optimal ambiguity graph operating on the boundary of the
detection-association trade-off.

We term the first subproblem the \emph{ambiguity-aware beamforming} problem. We
show that the ambiguity-aware beamforming problem can be formulated as a
semidefinite optimization problem, which is efficiently solvable using
interior-point methods. Through analysis and several simulations, we demonstrate
two types of performance improvements due to the beamforming being ambiguity
aware: an improvement in \emph{transmit power gain} and an improvement in
\emph{target identifiability}. The transmit power gain increases detection
performance for a fixed number of targets. The improvement in target
identifiability increases the number of targets that can be tracked for a fixed
number of radar antennas. Through analysis and numerical simulations, we
demonstrate that both of these gains can be quite substantial.

To solve the second subproblem, we propose an \emph{ambiguity-aware
nearest-neighbor} association scheme. This association scheme is a
generalization of the well-known nearest-neighbor association scheme. Unlike
that scheme, it explicitly takes the structure of the beam pattern (captured by
the ambiguity graph) into account. The proposed association scheme has a cubic
worst-case computational complexity in the number of targets being tracked.

Finally, we consider the third subproblem of choosing a Pareto-optimal ambiguity
graph operating on the boundary of the detection-association trade-off. This
problem turns out to be a computationally difficult combinatorial optimization
problem.  We therefore propose a design heuristic and show through simulations
that it can yield near-optimal performance.

\subsection{Related Work}
\label{sec:intro_related}

Multi-antenna radar systems can broadly be classified into two categories:
phased arrays~\cite[Chapter~3]{richards14}, in which phase-shifted versions of
the same signal are emitted from the antennas, and MIMO radar~\cite{listoica09},
in which general signals can be emitted from the antennas. The increased
adaptability of MIMO radar offers significant advantages over phased-array
radar, including increased target identifiability and flexibility in transmit beam
pattern design~\cite{listoica09}.

A number of prior works have studied the problem of designing transmit beam
patterns for MIMO radar systems. These methods vary in the amount of prior
target information, generally obtained from the track filters at the receiver,
they utilize. Some methods do not use any prior target information and focus on
creating beam patterns with wide coverage~\cite{yu16}. 

A large body of literature creates beam patterns adapted to the tracked targets'
azimuth angles. One such approach designs the beam pattern using a two-step
process, which first selects a cross-correlation matrix to achieve a desired
beam pattern and then chooses signals having that correlation~\cite{fuhrmann08}.
Another approach decouples the spatial and temporal dimensions of the
beamforming design~\cite{friedlander12, lipor14}. Yet another class of methods
adopts a beam-space approach to beamformer
design~\cite{fuhrmann10,li15b,hassanien11,khabbazibasmenj14,stoica07}, which
aims to combine the benefits of phased-array and MIMO radar.  Beamformer designs
that are robust to mismatches between the presumed and actual beamforming
vectors are studied in~\cite{zhang16}. In the automotive context, beamforming
has also been studied from the perspective of interference
avoidance~\cite{bechter16}. 

A few recent works~\cite{sharaga15,huleihel13} have addressed the question of
adapting beam patterns for improved detection. Some works focus on adapting
waveforms to account for change in clutter in the radar
scene~\cite{li09,friedlander06,friedlander07}, while other works focus on
adaptive beamforming to improve tracking~\cite{leshem07,sharaga15,tabrikian13}.

The different methods described above design beamforming patterns to maximize
detection accuracy and are not directly concerned with the resulting association
problem. In contrast, in this paper we explicitly take the data association
problem into account while designing the beam pattern, and we show that there is
in fact a trade-off between the detection and the association performance.

\subsection{Organization}
\label{sec:intro_organization}

The remainder of this paper is organized as follows. Section~\ref{sec:problem}
introduces the problem setting. Section~\ref{sec:beam} shows how to solve the
ambiguity-aware beamforming problem. Ambiguity-aware association schemes and
how to choose a Pareto-optimal ambiguity graph are discussed in
Section~\ref{sec:graph}.  Section~\ref{sec:conclusion} contains discussions and
concluding remarks.

\section{Problem Setting}
\label{sec:problem}

We consider a monostatic radar system with $N$ antennas. Denote by
$a_n^\ast(\theta)\in\C$ and $b_n(\theta)\in\C$ the transmit and receive gains of
antenna $n$ at azimuth angle $\theta$. Assume we emit the complex,
baseband-equivalent, vector-valued signal $\bm{s}(t)\in\C^N$ with $t\in[0,T)$
from the antenna array.  The corresponding (noiseless, baseband-equivalent)
received signal after reflection from a point target with unit radar
cross-section at range $\tau c/2$, Doppler shift $\omega$, and azimuth angle
$\theta$ is 
\begin{equation*}
    \bm{x}(t; \tau, \omega, \theta) 
    \defeq \bm{b}(\theta) \bm{a}^\dagger(\theta) \bm{s}(t-\tau)\exp(j\omega t).
\end{equation*}
Note that here and in the following we use lowercase bold font for vectors and
uppercase bold font for matrices. The vector $\bm{a}(\theta)$ has components
$a_n(\theta)$ and similar for the vector $\bm{b}(\theta)$. The symbol
$\cdot^\dagger$ denotes the conjugate transpose. Finally, $c$ denotes the speed
of light. For general radar cross-section $h\in\C$, the reflected signal is
$h\bm{x}(t; \tau, \omega, \theta)$. Observe that the radar cross-section $h$
includes the path loss.

For $K$ targets with parameters $(\tau_k, \omega_k, \theta_k)$ and radar
cross-section $h_k$ for $k\in \{1, 2, \dots, K\}$, the received signal is then
\begin{equation*}
    \bm{y}(t) \defeq \sum_{k=1}^K h_k \bm{x}(t; \tau_k, \omega_k, \theta_k) + \bm{z}(t),
\end{equation*}
where $\bm{z}(t)$ is Gaussian receiver noise.

We aim to track the parameters of these $K$ targets from the received signal
$\bm{y}(t)$. The standard receiver architecture uses a matched filter bank, with
each filter tuned to one parameter triple.  The output of the matched
filter tuned to $(\tau, \omega, \theta)$ is
\begin{align}
    \label{eq:rdef}
    r(\tau, \omega, \theta)
    & \defeq \sum_{n=1}^N \int_{t=\tau}^{T+\tau} y_n(t) x_n^\ast(t; \tau, \omega, \theta)\,dt \notag\\
    & = \sum_{n=1}^N \int_{t=\tau}^{T+\tau} 
    y_n(t) \bm{s}^\dagger(t-\tau)\exp(-j\omega t)\,dt\,\bm{a}(\theta) b_n^\ast(\theta).\hspace{0.4cm}
\end{align}
 
Following~\cite{friedlander12}, we make throughout the remainder of this paper
the assumption that the transmitted signal has the form
\begin{equation*}
    \bm{s}(t) \defeq \bm{W} \tilde{\bm{s}}(t)
\end{equation*}
for some matrix $\bm{W}\in\C^{N\times N}$ satisfying $\tr(\bm{W}\bm{W}^\dagger)
= 1$ and for some $\tilde{\bm{s}}(t)\in\C^{N}$ with support $t\in[0,T)$ and with
approximate bandwidth $B$. Here, $\tilde{\bm{s}}(t)$ is assumed to have good
cross-correlation properties, meaning that it has matrix-valued ambiguity function
$\bm{\chi}(\Delta\tau, \Delta\omega) \in \C^{N\times N}$ satisfying 
\begin{align}
    \label{eq:corr}
    \bm{\chi}(\Delta\tau, \Delta\omega) 
    & \defeq \int_{t=\max\{\Delta\tau,0\}}^{T+\min\{\Delta\tau,0\}}
    \tilde{\bm{s}}(t)\tilde{\bm{s}}^\dagger(t-\Delta\tau)\exp(-j\Delta\omega t)\,dt \notag\\
    & \approx \ind_{(0,0)}(\Delta\tau, \Delta\omega) \bm{I},
\end{align}
where
\begin{equation*}
    \ind_{(0,0)}(\Delta\tau, \Delta\omega) \defeq
    \begin{cases}
        1, & \text{if $(\Delta\tau, \Delta\omega) = (0, 0)$}, \\
        0, & \text{otherwise}.
    \end{cases}
\end{equation*}
The temporal support $T$ and the bandwidth $B$ of the waveforms place limits of
order $1/B$ on the resolution of $\Delta\tau$ and of order $1/T$ on the
resolution of $\Delta\omega$. The identity~\eqref{eq:corr} holds up to those
resolution limits. See Appendix~\ref{sec:app_corr} for a formal definition of
these limits. In the same appendix, we also describe and analyze a construction
for waveforms having property~\eqref{eq:corr} for large enough values of
time-bandwidth product $BT$. For simplicity, we make in the remainder of the
paper the assumption that~\eqref{eq:corr} holds with equality, i.e., that the
ambiguity function is ideal.

Assuming an ideal ambiguity function, the matched filter
output~\eqref{eq:rdef} can be simplified after some algebra to
\begin{align}
    \label{eq:r}
    r(\tau, \omega, \theta)
    & = \sum_{k=1}^K h_k \exp\bigl(-j(\omega-\omega_k) \tau_k\bigr)
    1_{(0,0)}(\tau-\tau_k, \omega-\omega_k)
    \bm{b}^\dagger(\theta)\bm{b}(\theta_k)
    \bm{a}^\dagger(\theta_k) \bm{R} \bm{a}(\theta) 
    + \tilde{z}(\tau, \omega, \theta), 
\end{align}
where 
\begin{equation}
    \label{eq:R}
    \bm{R} \defeq \bm{W}\bm{W}^\dagger
\end{equation}
is the \emph{beamforming matrix} and $\tilde{z}(\tau, \omega, \theta)$ is the
filtered receiver noise. Observe that $\bm{R}$ satisfies $\tr(\bm{R}) = \tr(
\bm{W}\bm{W}^\dagger) = 1$.

A common procedure for tracking the parameters $(\tau_k, \omega_k, \theta_k)$
for the $K$ targets consists of two steps, a \emph{detection} step followed by
an \emph{association} step. In the first step, we detect the targets. This is
done by thresholding the matched filter outputs $r(\tau, \omega, \theta)$. I.e.,
we find all $(\hat{\tau}, \hat{\omega}, \hat{\theta})$ such that
$\abs{r(\hat{\tau}, \hat{\omega}, \hat{\theta})}$ is strictly above some
threshold. In the second step, we associate each such detection $(\hat{\tau},
\hat{\omega}, \hat{\theta})$ with a target track using an association rule. 

Assume that we have prior information about the parameters $(\tau_k, \omega_k,
\theta_k)$. This prior is typically derived from the output of the prediction
stage of a filter tracking these targets. It can be used to define a
\emph{gate}, i.e., a region of parameter space in which we expect the true
parameters for a fixed target to fall. Similar to other works in the
literature~\cite{friedlander12}, we make for now the simplifying assumption that
the prior information on the azimuth angle is precise, so that $\theta_k$ is in
effect known a priori.  Section~\ref{sec:conclusion} discusses how this
assumption can be relaxed. With this assumption, the gate $\mc{S}_k$ for target
$k$ is a subset of $\R^2$ in which we expect the parameter tuple $(\tau_k,
\omega_k)$ to fall.

We now formalize a somewhat stylized association rule that we will use
throughout the paper: For each target track
$k\in\{1,2,\dots,K\}$, find all detections of the form $(\hat{\tau},
\hat{\omega}, \hat{\theta})$ with $(\hat{\tau},\hat{\omega})\in\mc{S}_k$, and
$\hat{\theta} = \theta_k$ (which is known a priori by assumption). If there is a
single such detection, then we associate it with the track for target $k$. If
there are zero or more than one such detections, then we declare an association
error. In this association rule, the priors are used to gate the detections.
Further, the detection-to-track association is performed only if this gating
removes all association ambiguity. While this association rule is quite simple,
it includes several standard rules as special cases.

\begin{example}[name=Rectangular Gates, label=eg:gates_rect]
    From the prior, we can derive lower bounds $(\tau_k^-, \omega_k^-)$ and
    upper bounds $(\tau_k^+, \omega_k^+)$ for the parameter tuple $(\tau_k,
    \omega_k)$. For example, $\tau_k^+$ could be chosen as the prior mean of
    $\tau_k$ plus three standard deviations (and similarly for the other
    bounds). We can then construct the standard rectangular gating set $\mc{S}_k
    \defeq [\tau_k^-, \tau_k^+]\times[\omega_k^-, \omega_k^+]$.
\end{example}

\begin{example}[name=Nearest-Neighbor Gates, label=eg:gates_greedy]
    \begin{figure}
        \centering 
        \includegraphics{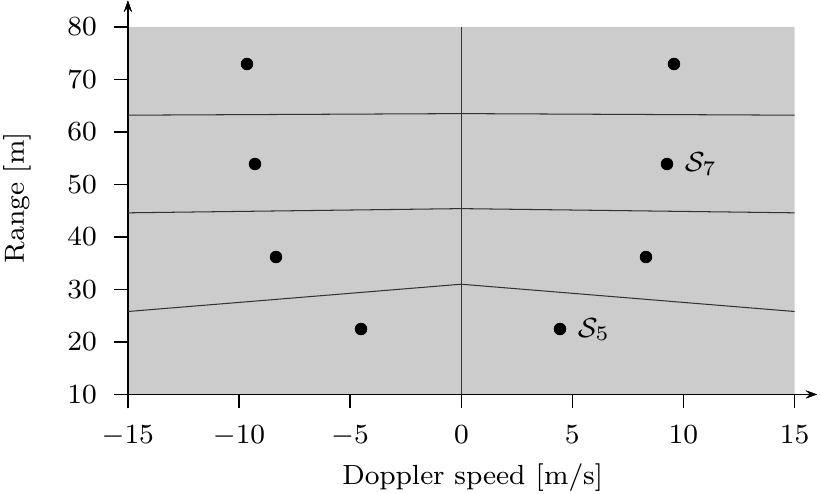}

        \caption{Illustration of nearest-neighbor gating sets $\{\mc{S}_k\}$.
        The dots show the expected target parameters from the priors. 
        The figure assumes equal-variance Gaussian priors, so that the resulting
        gating sets form a Voronoi partition.}

        \label{fig:gates_greedy}
    \end{figure}

    Let $f_k(\tau,\omega)$ be the prior density on the parameters
    $(\tau_k, \omega_k)$ of target $k$. Set
    \begin{equation*}
        \mc{S}_k \defeq 
        \bigl\{
            (\tau,\omega): f_k(\tau,\omega) > f_{k'}(\tau,\omega) \text{ for all
            $k'\neq k$}
        \bigr\}.
    \end{equation*}
    See Fig.~\ref{fig:gates_greedy} for an illustration.  The resulting gating
    sets partition the space $\R^2$. For Gaussian priors with equal variance,
    the resulting partition is a Voronoi partition. The corresponding association
    rule assigns each detection to the track it most likely came from. This is
    similar to the standard nearest-neighbor or greedy association rule.
\end{example}

For future reference, we define 
\begin{equation}
    \label{eq:Ek}
    \mc{E}_k \defeq \bigl\{k' \in \{1, 2, \dots, K\}\setminus\{k\}: \mc{S}_{k'}
    \cap \mc{S}_k \neq \emptyset\bigr\}
\end{equation}
as the targets $k'$ that are a-priori ambiguous with respect to target
$k$.

To understand the interaction between data association and beamforming, consider
initially a scenario with $K=2$ targets with known azimuth angles $\theta_1$ and
$\theta_2$. Assume the matched filter outputs $\abs{r(\tau_1, \omega_1,
\theta_1)}$, $\abs{r(\tau_1, \omega_1, \theta_2)}$, $\abs{r(\tau_2, \omega_2,
\theta_1)}$, $\abs{r(\tau_2, \omega_2, \theta_2)}$ are all above the detection
threshold. Without further knowledge we cannot unambiguously solve the
association problem.  Indeed, $(\tau_1, \omega_1)$ could be correctly associated
with target one and $(\tau_2, \omega_2)$ correctly with target two. However,
alternatively, $(\tau_1, \omega_1)$ could be incorrectly associated with target
two and $(\tau_2, \omega_2)$ incorrectly with target one. Without additional
information, both associations are equally valid from the radar receiver's point
of view.

This ambiguity can be resolved in two ways. First, through additional prior
information about $\tau_k$ and $\omega_k$. If the gates $\mc{S}_1$ and $\mc{S}_2$
do not intersect (or, equivalently, if $2\notin\mc{E}_1$), then one of the two
data associations is invalidated by the prior.

Second, by designing the transmitted signals to ensure that two of the matched
filter outputs are less than the detection threshold, say $\delta$. I.e.,
\begin{align*}
    \abs{r(\tau_1, \omega_1, \theta_2)} & \leq \delta, \\
    \abs{r(\tau_2, \omega_2, \theta_1)} & \leq \delta.
\end{align*}
In this case, one of the two data associations is invalidated by the
observations.

Consider next the general case with $K$ targets.  Assume for the moment that the
received signal is noiseless, and set the detection threshold to zero. Under
this assumption and using~\eqref{eq:r}, the matched filter output $r(\hat{\tau},
\hat{\omega}, \hat{\theta})$ is nonzero, and hence a potential target detected,
if and only if $(\hat{\tau}, \hat{\omega}) = (\tau_{k}, \omega_{k})$ for some
$k\in\{1, 2, \dots, K\}$ satisfying
\begin{equation*}
    \abs{ h_k \bm{b}^\dagger(\hat{\theta})\bm{b}(\theta_k)
    \bm{a}^\dagger(\theta_k) \bm{R} \bm{a}(\hat{\theta}) }
    \neq 0.
\end{equation*}
Consider the resulting association problem. Our association rule (defined just
above Example~\ref{eg:gates_rect}) will assign all detections to tracks without
declaring an error if, for every $k$, there is exactly one tuple $(\hat{\tau},
\hat{\omega})$ inside the gating set $\mc{S}_k$ for target $k$ such that
$r(\hat{\tau}, \hat{\omega}, \theta_k)$ is nonzero.

From the above discussion, the association problem has a unique solution if
the following two sufficient conditions hold. First, for every $k\in\{1, 2,
\dots, K\}$,
\begin{equation*}
    \abs{h_k} \norm{\bm{b}(\theta_k)}^2 \bm{a}^\dagger(\theta_k) \bm{R} \bm{a}(\theta_k) > 0,
\end{equation*}
so that the correct target is detected. Second, for every $k\in\{1, 2, \dots,
K\}$ and every $k' \in\mc{E}_k$,
\begin{equation*}
    h_k \bm{b}^\dagger(\theta_{k'})\bm{b}(\theta_k) \bm{a}^\dagger(\theta_k) \bm{R} \bm{a}(\theta_{k'}) = 0,
\end{equation*}
so that hard to disambiguate targets are not simultaneously illuminated. This
second condition ensures that the association is unambiguous. Crucially, these
conditions depend solely on the prior knowledge of the target
parameters (i.e., the gates $\mc{S}_k$ and the azimuths $\theta_k$).

To focus on the essential aspect of the problem, we enforce in the following the
slightly stronger condition that
\begin{equation}
    \label{eq:detectability}
    \bm{a}^\dagger(\theta_k) \bm{R} \bm{a}(\theta_k) > 0,
\end{equation}
for every $k$ and that
\begin{equation}
    \label{eq:unambiguity}
    \bm{a}^\dagger(\theta_k) \bm{R} \bm{a}(\theta_{k'}) = 0
\end{equation}
for every $k$ and $k' \in\mc{E}_k$.

To guard against receiver noise, we further want the
transmit power gaing~\eqref{eq:detectability} to be as large as possible for 
all targets. In fact, for fixed detection threshold, we can upper bound the 
probability of detection error as a decreasing function of the left-hand side
of~\eqref{eq:detectability} for the worst target $k$ (see,
e.g.,~\cite[Chapter~6.3]{richards14}). The optimal \emph{ambiguity-aware
beamforming problem} is therefore\footnote{To minimize the aforementioned
probability of detection error, the term $\abs{h_k}\norm{\bm{b}(\theta_k)}^2$ 
should be added to the objective function of~\eqref{eq:problem}. We prefer to 
drop this term here to simplify the notation, since it does not change the 
nature of the problem as discussed above. Further, the hard zero-forcing
constraint~\eqref{eq:unambiguity} could be replaced by a weaker constraint
that the interfering signal contribution is below the noise floor, as is
discussed in Section~\ref{sec:conclusion}.}
\begin{maxi}
    {%
        \bm{R}\in\C^{N\times N}%
    }
    {%
        \min_{k\in\{1,2,\dots,K\}} \bm{a}^\dagger(\theta_k) \bm{R} \bm{a}(\theta_k)%
    }
    {\label{eq:problem}}{}
    \addConstraint{\bm{a}^\dagger(\theta_k) \bm{R} \bm{a}(\theta_{k'})}{= 0, \quad}{\text{$\forall k$ and $\forall k'\in\mc{E}_k$}}
    \addConstraint{\tr(\bm{R})}{= 1}{}
    \addConstraint{\bm{R}}{ \succeq 0}{}
\end{maxi}
Here $\bm{R} \succeq 0$ denotes that the matrix $\bm{R}$ is positive
semidefinite (which implies that it is Hermitian).

The structure of the ambiguity-aware beamforming problem can be conveniently
described by a graph with vertices $\{1, 2, \dots, K\}$ (one for each target)
and an edge between vertex $k$ and $k'$ if $k'\in\mc{E}_k$.  Since
$k'\in\mc{E}_k$ if and only if $k\in\mc{E}_{k'}$ by construction, this graph is
undirected. We refer to this graph as the \emph{ambiguity graph} in the
following. The next example introduces a special case of this graph that will be
used to illustrate the results throughout the paper.

\begin{example}[name=Ambiguity Graph, label=eg:graph]

    \begin{figure}
        \centering 
        \subfigure[Vehicular scenario\label{fig:grapha}]{\hspace{0.5cm}\includegraphics[width=7.6cm]{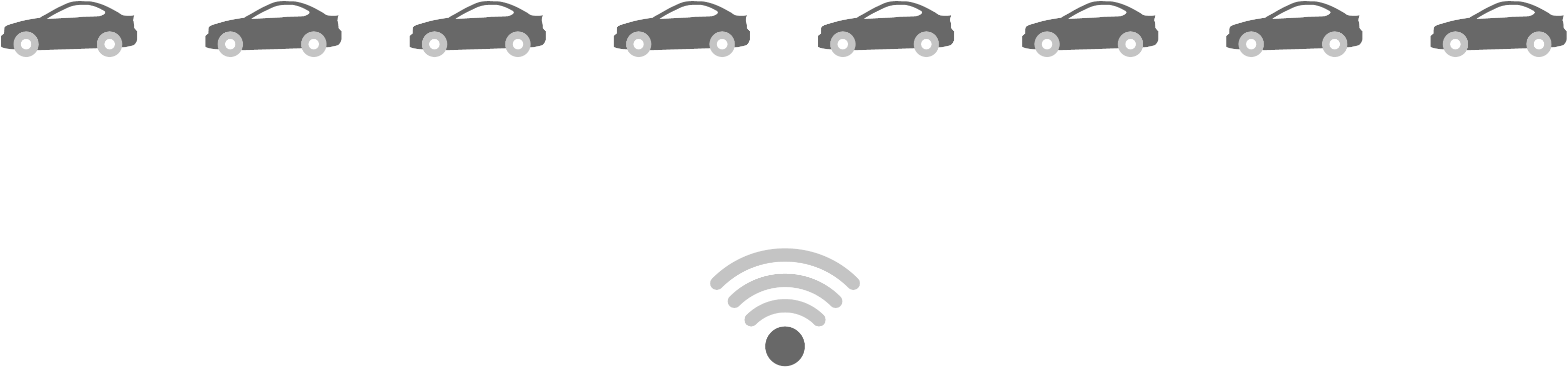}}\\
        \subfigure[Doppler-range plane\label{fig:graphb}]{\hspace{-0.5cm}\includegraphics{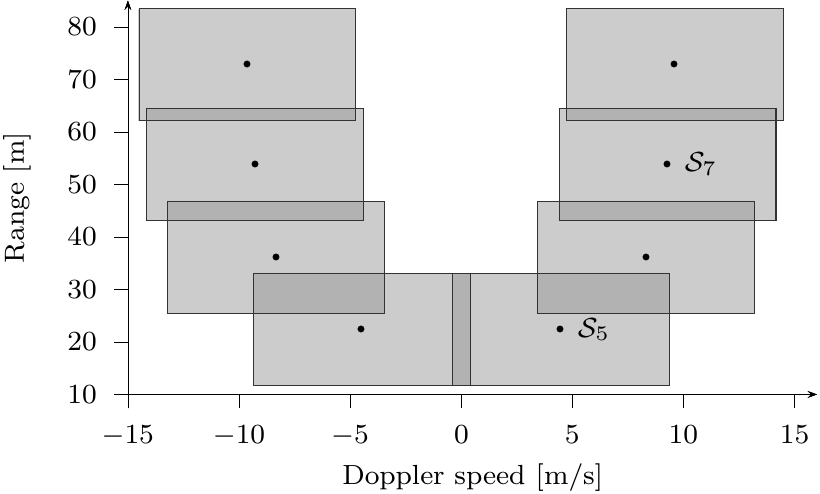}}\\
        \subfigure[Ambiguity graph\label{fig:graphc}]{\hspace{0.5cm}\includegraphics{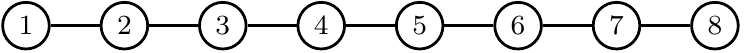}}\\

        \caption{Vehicular scenario, resulting parameter ambiguity in the
        Doppler-range plane, and corresponding ambiguity graph.}

        \label{fig:graph}
    \end{figure}

    Consider a vehicular scenario with a radar at a crossroad on which several
    cars are driving in the same direction. This scenario is depicted in
    Fig.~\ref{fig:grapha}. The corresponding expected values of the target
    parameters $\tau_k$ and $\omega_k$ are shown in Fig.~\ref{fig:graphb} by
    black dots together with rectangular gates $\mc{S}_k \defeq [\tau_k^-,
    \tau_k^+]\times[\omega_k^-, \omega_k^+]$ as defined in
    Example~\ref{eg:gates_rect} indicated in gray.  Observe from
    Fig.~\ref{fig:graphb} that only the gating sets of neighboring vehicles
    intersect. The resulting ambiguity graph is depicted in
    Fig.~\ref{fig:graphc}.
\end{example}

\begin{remark}
    If we have no prior information about the parameters $\tau_k$ and
    $\omega_k$, then all targets are a-priori ambiguous. The resulting ambiguity
    graph is therefore complete (i.e., has and edge between every pair of
    distinct targets). In this case, the ambiguity-aware beamforming problem
    essentially reduces to the setting in~\cite{friedlander12}. However, for
    partial ambiguity graphs, like the one introduced in Example~\ref{eg:graph},
    the problem and its solution are quite different. This connection is
    discussed further in Example~\ref{eg:friedlander} in Section~\ref{sec:beam}.
\end{remark}

\section{Optimal Ambiguity-Aware Beamforming}
\label{sec:beam}

The ambiguity-aware beamforming problem defined in~\eqref{eq:problem} can be
solved efficiently as follows.  Note that the function $\bm{a}^\dagger(\theta_k)
\bm{R} \bm{a}(\theta_k)$ is linear in $\bm{R}$ and hence, in particular,
concave. Since the minimum of concave functions is again concave, this implies
that the objective function of~\eqref{eq:problem} is concave in $\bm{R}$. The
conditions $\bm{a}^\dagger(\theta_k) \bm{R} \bm{a}(\theta_{k'}) = 0$ and
$\tr(\bm{R}) = 1$ are linear in $\bm{R}$. Further, we have the positive
semidefiniteness constraint $\bm{R} \succeq 0$, which describes a convex set.
Thus, we are dealing with a convex problem, which can actually be rewritten as a
semidefinite program by introducing a slack variable to handle the minimization
over $k$. These programs can be solved efficiently using interior-point
methods~\cite{boyd04}.

We illustrate this approach with several examples.

\begin{example}[label=eg:3x3]

    \begin{figure}
        \centering 
        \subfigure[$\abs{\bm{a}^\dagger(\theta_1)\bm{R}\bm{a}(\theta)}$\label{fig:3x3i1}]{\includegraphics{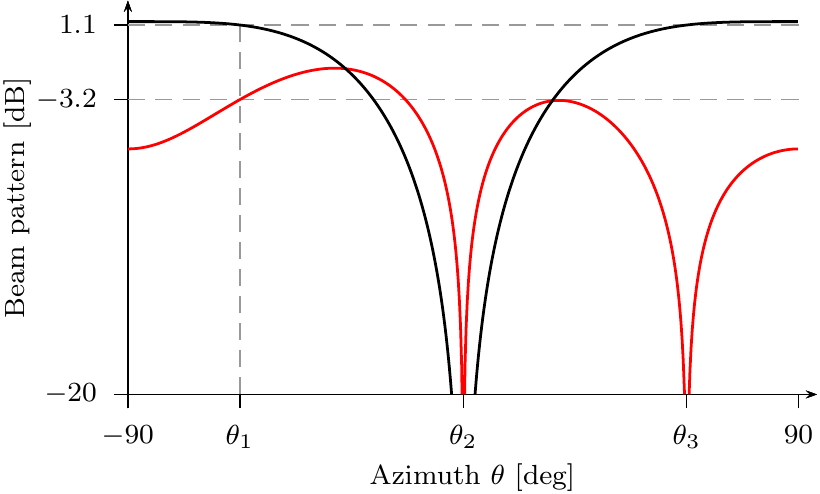}}\\
        \subfigure[$\abs{\bm{a}^\dagger(\theta_2)\bm{R}\bm{a}(\theta)}$\label{fig:3x3i2}]{\includegraphics{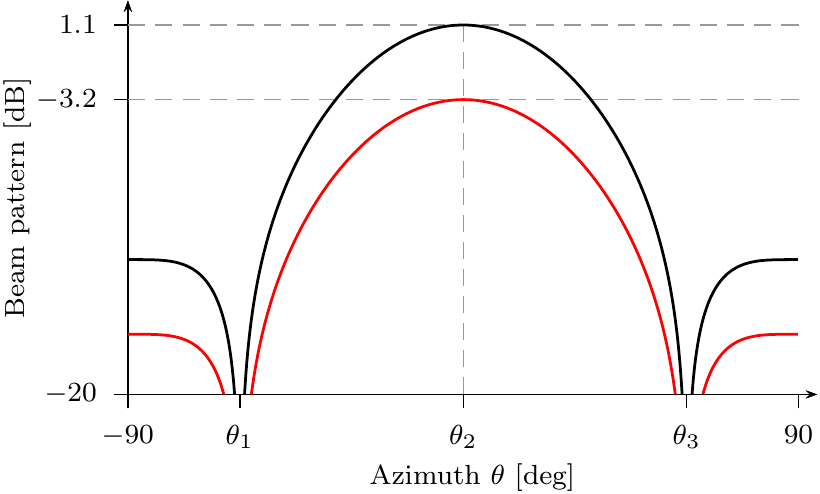}}\\
        \subfigure[$\abs{\bm{a}^\dagger(\theta_3)\bm{R}\bm{a}(\theta)}$\label{fig:3x3i3}]{\includegraphics{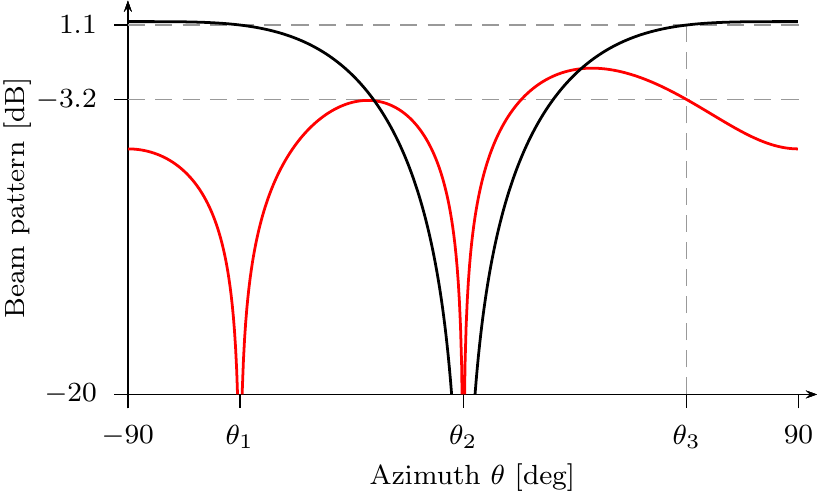}}\\

        \caption{Optimal transmit beam patterns for two scenarios with $N = 3$ 
            antennas and $K = 3$ targets. The red/gray curve
            (\textcolor{red}{\protect\rule[2pt]{4pt}{1pt}})
            is for the scenario of complete ambiguity.
            The black curve (\protect\rule[2pt]{4pt}{1pt})
            is for the scenario of ambiguity only between neighbors (i.e., not
        between $\theta_1$ and $\theta_3$).}
        \label{fig:3x3}
    \end{figure}
    
    Consider a scenario with $N = 3$ uniformly spaced antennas at
    half-wavelength separation and with $K = 3$ targets at azimuth angles
    $\theta_1 = \SI{-60}{\degree}$, $\theta_2 = \SI{0}{\degree}$, and $\theta_3
    = \SI{60}{\degree}$. The corresponding transmit antenna gains are
    \begin{equation}
        \label{eq:ula}
        a_n(\theta) \defeq \exp\bigl(j(n-1)\pi\sin(\theta)\bigr).
    \end{equation}

    Assume first that the ambiguity graph is complete (i.e., $\mc{E}_k \defeq
    \{1, 2, \dots, K\}\setminus\{k\}$ for each $k$ so that the ambiguity graph
    has an edge between every pair of targets). The corresponding optimal
    transmit beam pattern $\bm{a}^\dagger(\theta_k)\bm{R}\bm{a}(\theta)$ with
    $\bm{R}$ the solution to~\eqref{eq:problem} as a function of azimuth angle
    $\theta$ for fixed $\theta_k$ is shown in red/gray in Fig.~\ref{fig:3x3}.

    Assume next that the ambiguity graph contains only edges between neighboring
    targets (as introduced in Example~\ref{eg:graph}). The corresponding optimal
    transmit beam pattern is shown in black in Fig.~\ref{fig:3x3}.

    Comparing the two curves in Fig.~\ref{fig:3x3}, two key differences are
    apparent. First, the beam pattern for the complete ambiguity graph has an
    additional null in Figs.~\ref{fig:3x3i1} and~\ref{fig:3x3i3}. Second, by not
    having to enforce these nulls, the beam pattern for the partial ambiguity
    graph is able to increase the transmit power gain at the desired target by
    $\SI{4.3}{\deci\bel}$ (from $\SI{-3.2}{\deci\bel}$ to
    $\SI{1.1}{\deci\bel}$).
\end{example}

As the previous example shows, taking the reduced ambiguity between the targets
(captured by the ambiguity graph being incomplete) into account can improve the
transmit power gain. This improvement is explored further in the next example.

\begin{example}[name=Transmit Power Gain, label=eg:snr]

    \begin{figure}
        \centering 
        \includegraphics{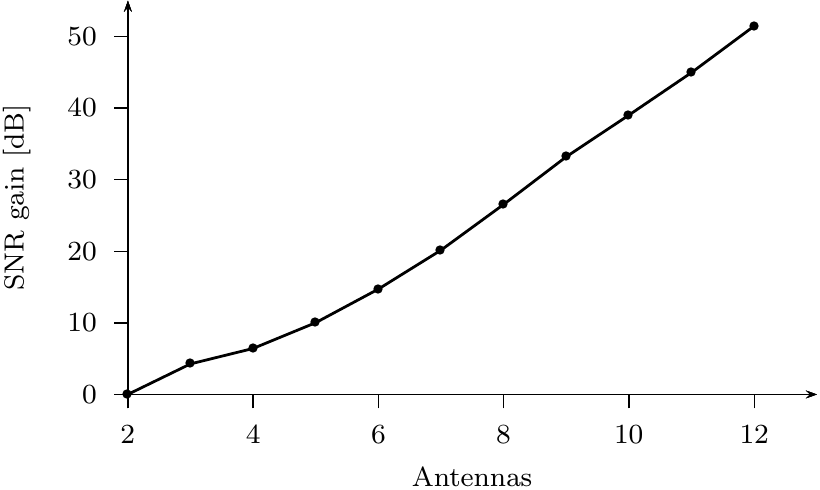}

        \caption{Improvement in transmit power gain of beamforming for partial
        ambiguity over complete ambiguity as a function of number of antennas
        $N$.}
        \label{fig:snr}
    \end{figure}

    Consider $N$ uniformly spaced antennas at half-wavelength separation and
    with $K = N$ targets. Target $k\in\{1, 2, \dots, K\}$ has azimuth angle
    $\theta_k \defeq \bigl( (2k-1)/K -1\bigr)\SI{90}{\degree}$. In words, the
    targets are uniformly spaced in azimuth. For $N=K=3$, this reduces to the
    setting in Example~\ref{eg:3x3}.

    We compare the optimal beamforming matrices designed according
    to~\eqref{eq:problem} for a complete ambiguity graph and for a partially
    connected graph with edges only between neighboring targets $k$ and $k+1$
    for all $k\in\{1, 2, \dots, K-1\}$ (see again Example~\ref{eg:graph}). The
    transmit power gain ratio, i.e., the ratio of the quantity
    $\min_{k\in\{1,2,\dots,K\}} \bm{a}^\dagger(\theta_k) \bm{R}
    \bm{a}(\theta_k)$ between the two scenarios is shown in Fig.~\ref{fig:snr}
    as a function of the number of transmit antennas $N$. As is clear from the
    figure, explicitly taking the target ambiguity into account can yield a
    significant improvement in transmit power gain that increases with the
    problem size.
\end{example}

The last example indicates that ambiguity-aware beamforming can substantially
improve the transmit power gain. In other words, for the same number of targets,
we can expect better detection performance. As we see next, ambiguity-aware
beamforming can also improve target identifiability, i.e., for the same number
of antennas, we can track more targets.

\begin{example}[name=Target Identifiability, label=eg:ident]

    \begin{figure}
        \centering 
        \includegraphics{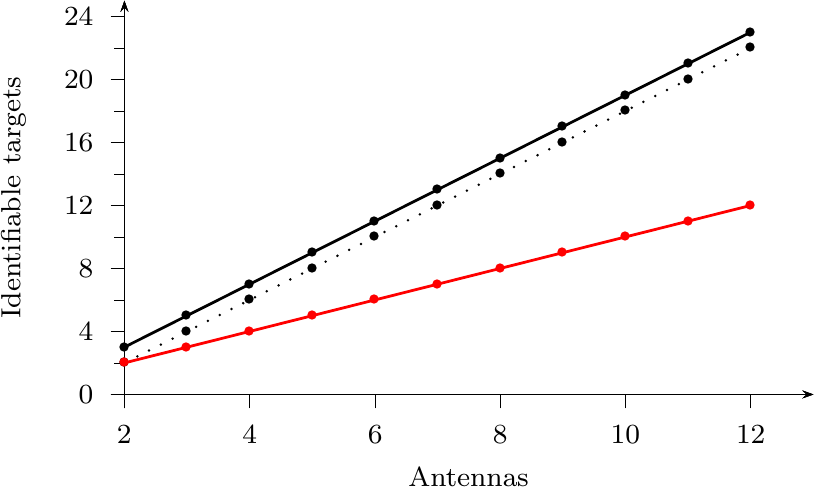}

        \caption{Number of identifiable targets $K^\star$ as a function of number of
            antennas $N$ for complete ambiguity (\textcolor{red}{\protect\rule[2pt]{4pt}{1pt}})
            and partial ambiguity (exact numerical value \protect\rule[2pt]{4pt}{1pt} and
            analytical lower bound \protect\parbox{8pt}{\dotfill}).}

        \label{fig:ident}
    \end{figure}

    Consider again the scenario of Example~\ref{eg:snr} with $N$ uniformly
    spaced antennas and $K$ uniformly spaced targets. However, this time we
    allow $K$ and $N$ to differ. We are interested in the maximum number of
    identifiable targets, which we denote by $K^\star$ and define formally as
    the largest $K$ for which the optimization problem~\eqref{eq:problem} has a
    solution.
    
    We compare again the complete ambiguity graph and the partially connected
    ambiguity graph with edges only between neighboring targets as introduced in
    Example~\ref{eg:graph}. For the complete graph we show analytically in
    Appendix~\ref{sec:app_ident} that $K^\star = N$ (see the solid red/gray
    curve in Fig.~\ref{fig:ident}). In contrast, the solid black curve in
    Fig.~\ref{fig:ident} shows the numerically computed value of $K^\star$ for
    the partial graph. The figure suggests that the number of identifiable
    targets $K^\star$ is $2N-1$, i.e., almost doubled.  Thus, explicitly taking
    the target ambiguity into account can also yield a significant improvement
    in target identifiability that again increases with the problem size.
\end{example}

We next explore the structure of the solution to the ambiguity-aware
beamforming problem. We start by introducing some additional notation. 

Denote by $\bm{U}_k\in\C^{N\times(N-\card{\mc{E}_k})}$ a matrix whose columns
are an orthonormal basis for the orthogonal complement of the subspace spanned
by $\{\bm{a}(\theta_{k'})\}_{k'\in\mc{E}_k}$. In other words, $\bm{U}_k^\dagger
\bm{U}_k = \bm{I}$, and $\bm{U}_k^\dagger\bm{a}(\theta_{k'}) = \bm{0}$ for all
$k'\in\mc{E}_k$.  Further, denote by $\bm{V}_k\in\C^{N\times(N-1)}$ a matrix
whose columns are an orthonormal basis for the orthogonal complement of the
subspace spanned by $\bm{a}(\theta_k)$. In other words, $\bm{V}_k^\dagger
\bm{V}_k = \bm{I}$, and $\bm{V}_k^\dagger\bm{a}(\theta_{k}) = \bm{0}$. The
choice of $\bm{U}_k$ and $\bm{V}_k$ is not unique; any such matrices will work
for our purposes.

Consider then the matrix
\begin{equation}
    \label{eq:RFG}
    \bm{R}(\bm{F}_k, \bm{G}_k)
    \defeq \bm{U}_k \bm{F}_k \bm{U}_k^\dagger + \bm{V}_k \bm{G}_k \bm{V}_k^\dagger,
\end{equation}
where $\bm{F}_k\in\C^{(N-\card{\mc{E}_k})\times (N-\card{\mc{E}_k})}$ and
$\bm{G}_k\in\C^{(N-1)\times (N-1)}$ are arbitrary positive semidefinite
matrices. Note that $\bm{R}(\bm{F}_k, \bm{G}_k)$ is positive semidefinite. 
For any $k$ and $k'\in\mc{E}_k$,
\begin{align*}
    \bm{a}^\dagger(\theta_k)\bm{R}(\bm{F}_k, \bm{G}_k)\bm{a}(\theta_{k'})
    & = \bm{a}^\dagger(\theta_k)\bm{U}_k \bm{F}_k \bm{U}_k^\dagger\bm{a}(\theta_{k'})
    + \bm{a}^\dagger(\theta_k)\bm{V}_k \bm{G}_k \bm{V}_k^\dagger\bm{a}(\theta_{k'}) \\
    & = \bm{a}^\dagger(\theta_k)\bm{U}_k \bm{F}_k \bm{0} 
    + \bm{0}^\dagger\bm{G}_k \bm{V}_k^\dagger\bm{a}(\theta_{k'}) \\
    & = 0,
\end{align*}
and for any $k$
\begin{equation*}
    \bm{a}^\dagger(\theta_k)\bm{R}(\bm{F}_k, \bm{G}_k)\bm{a}(\theta_k)
    = \bm{a}^\dagger(\theta_k)\bm{U}_k\bm{F}_k\bm{U}_k^\dagger\bm{a}(\theta_k).
\end{equation*}
Further,
\begin{align*}
    \tr\bigl(\bm{R}(\bm{F}_k, \bm{G}_k)\bigr) 
    & = \tr\bigl(\bm{U}_k \bm{F}_k \bm{U}_k^\dagger\bigr) 
    + \tr\bigl(\bm{V}_k \bm{G}_k \bm{V}_k^\dagger\bigr) \\
    & = \tr\bigl(\bm{U}_k^\dagger\bm{U}_k \bm{F}_k \bigr) 
    + \tr\bigl(\bm{V}_k^\dagger\bm{V}_k \bm{G}_k \bigr) \\
    & = \tr(\bm{F}_k) + \tr(\bm{G}_k).
\end{align*}

From this discussion, we see that one way to construct a beamforming matrix
$\bm{R}$ is to choose $\bm{F}_k$, $\bm{G}_k$ for $k\in\{1, 2, \dots, K\}$ such that
\begin{equation*}
    \bm{R} 
    = \bm{R}(\bm{F}_1, \bm{G}_1)
    = \dots
    = \bm{R}(\bm{F}_K, \bm{G}_K),
\end{equation*}
and so that the constraint $\tr(\bm{R}) = 1$ is satisfied.  Formally, a
(potentially suboptimal) solution to the ambiguity-aware beamforming
problem~\eqref{eq:problem} can be found by solving:
\begin{maxi}
    {%
        \bm{F}_k, \bm{G}_k\, \forall k
    }
    {%
        \min_{k\in\{1,2,\dots,K\}}
        \bm{a}^\dagger(\theta_k)\bm{U}_k\bm{F}_k\bm{U}_k^\dagger\bm{a}(\theta_k)
    }
    {\label{eq:problem2}}{}
    \addConstraint{\bm{R}(\bm{F}_k, \bm{G}_k)}{= \bm{R}(\bm{F}_1, \bm{G}_1), \quad}{\text{for all $k$}}
    \addConstraint{\tr(\bm{F}_1) + \tr(\bm{G}_1)}{= 1}{}
    \addConstraint{\bm{F}_k}{\succeq 0, \quad}{\text{for all $k$}}
    \addConstraint{\bm{G}_k}{\succeq 0, \quad}{\text{for all $k$}.}
\end{maxi}

This reformulation allows to obtain analytical insight into the structure of the
solution to the ambiguity-aware beamforming problem as the next two examples
illustrate.

\begin{example}[name=Comparison with~\cite{friedlander12}, label=eg:friedlander]
    Let $N = K$ and consider a scenario with a complete ambiguity graph. Then
    the matrix $\bm{U}_k\in\C^{N\times(N-\card{\mc{E}_k})}=\C^{N\times 1}$ is in
    fact simply a vector $\bm{u}_k$, and the matrix $\bm{F}_k$ is simply a
    positive scalar $f_k$. Thus,
    \begin{align*}
        \bm{R}(f_k, \bm{G}_k) 
        = f_k \bm{u}_k \bm{u}_k^\dagger + \bm{V}_k\bm{G}_k\bm{V}_k^\dagger.
    \end{align*}
    We need to ensure that
    \begin{equation*}
        \bm{R}(f_1, \bm{G}_1)
        = \dots
        = \bm{R}(f_K, \bm{G}_K).
    \end{equation*}
    Fix a target $k$ and consider another target $k'\neq k$. Since the
    ambiguity graph is complete, $\bm{u}_{k'}$ is orthogonal to
    $\bm{a}_k$. Since the columns of $\bm{V}_k$ span the orthogonal complement
    of the subspace spanned by $\bm{a}_k$, this implies that
    $\bm{u}_{k'}$ can be written as
    $\bm{V}_k\bm{\nu}_{k,k'}$ for some $\bm{\nu}_{k,k'}$. Set
    \begin{align*}
        \bm{G}_k \defeq \sum_{k'\neq k} f_{k'}
        \bm{\nu}_{k,k'}\bm{\nu}_{k,k'}^\dagger.
    \end{align*}
    Then
    \begin{align*}
        \bm{R}(f_k, \bm{G}_k) 
        & = f_k \bm{u}_k \bm{u}_k^\dagger + 
        \sum_{k'\neq k} f_{k'}
        \bm{V}_k\bm{\nu}_{k,k'}\bm{\nu}_{k,k'}^\dagger\bm{V}_k^\dagger \\
        & = \sum_{k'=1}^K f_{k'} \bm{u}_{k'} \bm{u}_{k'}^\dagger,
    \end{align*}
    which is identical for all $k$ as required. The values of $f_{k'}$
    can now be chosen to satisfy the trace constraint and to maximize the
    transmit power gain. This corresponds to the beamforming solution proposed
    in~\cite{friedlander12}.\footnote{To be precise,~\cite{friedlander12} sets
    $f_k \defeq f$ for all $k$ with $f$ chosen to satisfy the trace constraint.}
    Thus, for the special case of complete ambiguity graphs, the
    solution to~\eqref{eq:problem2} reduces to the one in~\cite{friedlander12}.
\end{example}

\begin{example}[name=Identifiability Gain, continues=eg:ident]
    We continue Example~\ref{eg:ident}, and assume for the moment that $K$ is
    even. As we had seen in Example~\ref{eg:friedlander}, for the case of complete
    ambiguity graphs, the beamforming problem and its solution reduce to the one
    in~\cite{friedlander12}. Further, as argued in Appendix~\ref{sec:app_ident},
    the largest number of identifiable targets $K^\star$ is equal to $N$ in
    this case.
    
    Consider then the case of a partial ambiguity graph with edges only between
    neighboring targets as defined in Example~\ref{eg:graph}. Construct the
    positive semidefinite matrix
    \begin{equation*}
        \bm{R} 
        \defeq \frac{1}{2N-K} \tilde{\bm{U}}_1\tilde{\bm{U}}_1^\dagger 
        + \frac{1}{2N-K}\tilde{\bm{U}}_2\tilde{\bm{U}}_2^\dagger, 
    \end{equation*}
    where the columns of $\tilde{\bm{U}}_1\in\C^{N\times (N-K/2)}$ are an
    orthonormal basis for the orthogonal complement of the space spanned by
    $\bm{a}(\theta_k)$ with $k$ odd, and where the columns of
    $\tilde{\bm{U}}_2\in\C^{N\times (N-K/2)}$ are an orthonormal basis for the
    orthogonal complement of the space spanned by $\bm{a}(\theta_k)$ with $k$
    even.  Observe that this is of a form similar to~\eqref{eq:RFG}. This
    construction ensures that
    $\bm{a}^\dagger(\theta_k)\bm{R}\bm{a}(\theta_{k+1}) = 0$, satisfying the
    zero-forcing constraint between neighboring targets. Further, it is easily
    seen that $\tr(\bm{R}) = 1$ as required.

    In general, this construction is possible as long as $N-K/2 \geq 1$. In
    other words, we can identify up to $2N-2$ targets, i.e., $K^\star \geq
    2N-2$. This is shown as the dashed black curve in Fig.~\ref{fig:ident}. It
    should be compared to the value of $K^\star$ of $2N-1$ obtained numerically
    shown as the solid black curve in Fig.~\ref{fig:ident}.
\end{example}

\section{Ambiguity-Aware Association and Pareto-Optimal Ambiguity Graphs}
\label{sec:graph}

So far, we have assumed that the ambiguity graph $G$ is given (defined
through~\eqref{eq:Ek} by the gates $\mc{S}_k\subset\R^2$).  In
Section~\ref{sec:beam}, we have seen how to solve the ambiguity-aware
beamforming problem~\eqref{eq:problem}, which maximizes the transmit power gain
$\bm{a}^\dagger(\theta_k)\bm{R}\bm{a}(\theta_k)$ for the worst-case target $k$
as a function of the ambiguity graph $G$. Denote by $P(G)$ the value of this
objective function for the optimal beamforming design.

We now turn to the problem of designing the ambiguity graph itself, or,
equivalently, the gating sets $\{\mc{S}_k\}_{k=1}^K$, for the given prior
distributions on the target parameters. Recall that our association rule is to
find all detections with received filter tuned to azimuth angle $\theta_k$
falling into the gate $\mc{S}_k$. If there is a single such detection, the rule
assigns it to track $k$. Otherwise, the rule declares an association error.
Correct association occurs if the association rule runs to completion (i.e.,
does not declare an association error) and assigns each detection to the correct
target.

Assume for the moment that the received signal is noiseless and a detection
threshold of zero is used. The probability of detection error is then equal to
zero. Further, making use of the zero-forcing constraint~\eqref{eq:unambiguity},
we can lower bound the probability of correct association by the quantity
\begin{align}
    \label{eq:pcbar}
    C(G)
    & \defeq
    \Pp\bigl( 
    (\tau_k, \omega_k) \in \mc{S}_k,
    (\tau_{k'}, \omega_{k'}) \notin \mc{S}_k 
    \text{ $\forall k$ and $\forall k'\notin\mc{E}_k$}
    \bigr) \notag\\
    & =
    \Pp\bigl( 
    (\tau_k, \omega_k) \in \mc{S}_k 
    \text{ for all $k$}
    \bigr),
\end{align}
where the second equality follows from the definition of $\mc{E}_k$
in~\eqref{eq:Ek}. The randomness in this expression is due to $\tau_k$ and
$\omega_k$ being random variables with distributions given by the prior.

Now consider noisy received signals. In this case, the probability of error can
be upper bounded by the sum of two terms: The first term is an upper bound on
the probability of detection error and is a decreasing function of the power
gain $P(G)$ as mentioned in Section~\ref{sec:beam}. The second term is $1-C(G)$,
capturing the performance of the association rule.

There is a trade-off between these two terms: Choosing $G$ to increase $P(G)$
causes to decrease $C(G)$. At one extreme consider the complete graph $G$
containing all $K(K-1)/2$ edges. For this graph, the received filter tuned to a
particular azimuth angle $\theta_k$ nulls out the interference of all other
targets $k'\neq k$. Therefore, the association problem becomes simple, and the
probability of correct association $C(G)$ is large. However, since we need to
create beams with $K(K-1)/2$ nulls, the optimization problem~\eqref{eq:problem}
is maximally constrained and hence the optimal transmit power gain $P(G)$ is
minimized.  At the other extreme, consider the empty graph $G$ containing no
edges. For this graph, the beams can be chosen without any nulling constraints.
Therefore, the transmit power gain $P(G)$ is maximized. However, the received
filter tuned to a particular azimuth angle $\theta_k$ contains interference from
all other targets $k'\neq k$. Therefore, the association problem is difficult,
and the probability of correct association $C(G)$ is small.

Formally, the detection-association trade-off is given by the set of all
Pareto-optimal\footnote{A pair $\bigl(P(G), C(G)\bigr)$ is Pareto optimal if no
other pair dominates it, meaning that there exists no other graph $G'$ such
that simultaneously $P(G') \geq P(G)$ and $C(G') \geq C(G)$, with at least
one of the two inequalities being strict.} pairs of the form $\bigl(P(G),
C(G)\bigr)$ parameterized by graphs $G$ with vertices $\{1, 2, \dots, K\}$. We
call an ambiguity graph $G$ optimal, if the corresponding pair $\bigl(P(G),
C(G)\bigr)$ is Pareto optimal. Since there are $2^{K(K-1)/2}$ possible graphs
with vertices $\{1, 2, \dots, K\}$, finding optimal ambiguity graphs represents
a computationally nontrivial combinatorial optimization problem.  We next
describe a design heuristic for constructing close to optimal ambiguity graphs. 

\begin{figure}
    \centering 
    \includegraphics{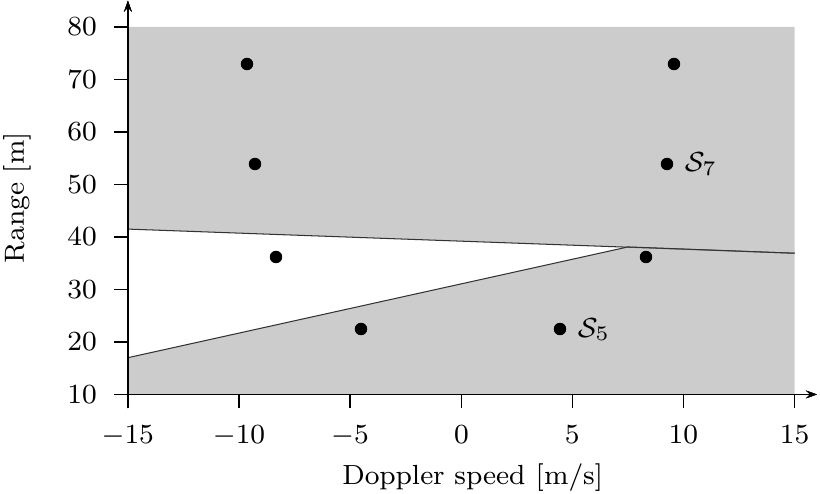}

    \caption{Illustration of ambiguity-aware nearest-neighbor gating sets
        $\mc{S}_k$ for the graph in Fig.~\ref{fig:graph}. For clarity, only two
        gating sets $\mc{S}_5$ and $\mc{S}_7$ are shown. Since the targets $5$
        and $7$ are not connected by an edge in the graph in
        Fig.~\ref{fig:graph}, the corresponding gating sets are disjoint.
        The figure assumes equal-variance Gaussian priors, resulting in convex
        polygonal regions.  However, the definition of $\mc{S}_k$ applies to
        arbitrary prior distributions, and the resulting regions are neither
        polygons nor convex in general. This figure can be compared to the
        standard nearest-neighbor gates shown in Fig.~\ref{fig:gates_greedy}.}

    \label{fig:scbar}
\end{figure}

For the expression~\eqref{eq:pcbar} to be well defined, we need to describe how
to choose the gating sets $\{\mc{S}_k\}_{k=1}^K$ corresponding to the ambiguity
graph $G$.  In order for the two to be consistent, we need $\mc{S}_k \cap
\mc{S}_{k'} = \emptyset$ for all $k$ and $k'\notin\mc{E}_k$ (see~\eqref{eq:Ek}).
We introduce here an extension of the nearest-neighbor gating sets seen in
Example~\ref{eg:gates_greedy}. Define
\begin{equation}
    \label{eq:skk}
    \mc{S}_{k,k'} \defeq
    \bigl\{
        (\tau,\omega) : f_k(\tau, \omega) > f_{k'}(\tau, \omega)
    \bigr\},
\end{equation}
where $f_k(\tau, \omega)$ is the probability density for the prior on $(\tau_k,
\omega_k)$. Further, define
\begin{equation*}
    \mc{S}_k \defeq
    \bigcap_{k'\notin\mc{E}_k} \mc{S}_{k,k'}
    =
    \Bigl\{
        (\tau,\omega) : f_k(\tau, \omega) > \max_{k'\notin\mc{E}_k} f_{k'}(\tau, \omega)
    \Bigr\}
\end{equation*}
as the set of $(\tau, \omega)$ pairs that are more likely to occur from 
target $k$ than from any of the non-neighboring targets $k'$
in the graph $G$. See Fig.~\ref{fig:scbar} for an illustration.
Clearly, $\mc{S}_k \cap \mc{S}_{k'} = \emptyset$ for all $k$ and
$k'\notin\mc{E}_k$, as required. 

\begin{algorithm}
    \caption{Ambiguity-aware nearest-neighbor association}
    \label{alg:nn}
    \begin{algorithmic}
        \ForAll {filter tracks $k$}
        \ForAll {detections $(\hat{\tau},\hat{\omega})$ with matched filter output $\abs{r(\hat{\tau}, \hat{\omega}, \theta_k)}$ larger than threshold}
        \If {$\argmax_{k'\notin\mc{E}_k} f_{k'}(\hat{\tau},\hat{\omega}) = k$}
        \State assign detection $(\hat{\tau}, \hat{\omega})$ to filter track $k$
        \EndIf
        \EndFor
        \EndFor
    \end{algorithmic}
\end{algorithm}
With this choice of gates $\mc{S}_k$, our generic association rule recalled at
the beginning of this section can be rewritten (after some algebra) more
compactly as in Algorithm~\ref{alg:nn}. Since the maximizing $k'$ in the
inner-most loop can be computed in $O(K)$ time, the entire algorithm has run
time $O(K^3)$.  We refer to this association rule as \emph{ambiguity-aware
nearest-neighbor association}.  We use this association rule for the evaluation
of $C(G)$.

Continuing from~\eqref{eq:pcbar}, we can now lower bound $C(G)$ using the
union bound as
\begin{align}
    \label{eq:pcg}
    C(G) 
    & \geq 
    \Pp\bigl(
    (\tau_k, \omega_k) \in \cap_{k'\notin\mc{E}_k} \mc{S}_{k,k'}
    \text{ for all $k$}
    \bigr) \notag\\
    & = 1-
    \Pp\bigl(
    (\tau_k, \omega_k) \in \cup_{k'\notin\mc{E}_k} \mc{S}_{k,k'}^{\msf{C}}
    \text{ for some $k$}
    \bigr) \notag\\
    & \geq 1-
    K^2\max_{(k,k'): k'\notin\mc{E}_k}
    \Pp\bigl(
    (\tau_k, \omega_k) \notin \mc{S}_{k,k'}
    \bigr).
\end{align}
This last lower bound is a function of only the most difficult to disambiguate
pair of targets $(k,k')$ not connected by an edge in the ambiguity graph $G$.

\begin{figure}
    \centering 
    \subfigure[$N=4$ antennas and $K=4$ targets.\label{fig:tradeoff4}]{\includegraphics{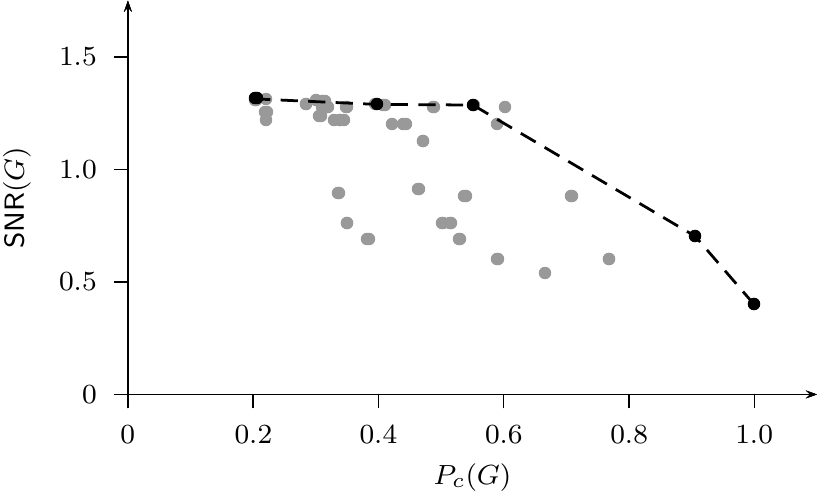}}%
    \hfill%
    \subfigure[$N=6$ antennas and $K=6$ targets.\label{fig:tradeoff6}]{\includegraphics{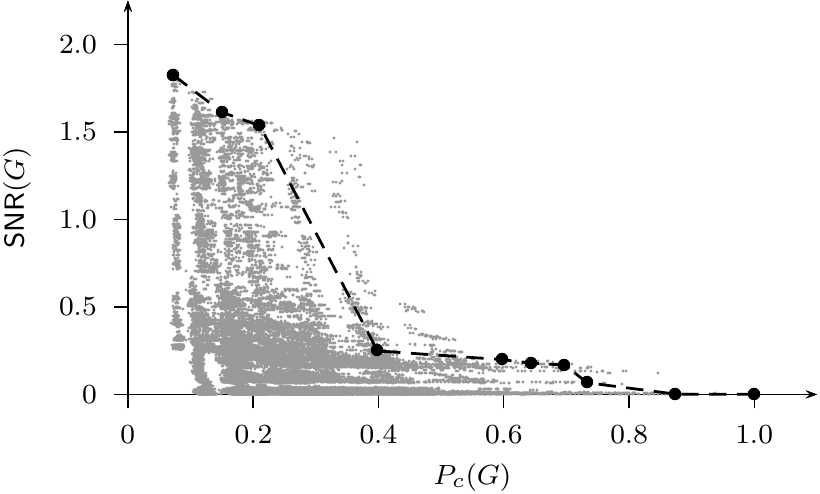}}

    \caption{Detection-association trade-off for two settings with targets
        positioned as in Fig.~\ref{fig:graph}. All values of $C(G)$ are
        numerically evaluated through Monte-Carlo integration. The black dots
        indicate the $\bigl(P(G_\gamma), C(G_\gamma)\bigr)$ pairs for different
        values of the threshold parameter $\gamma$. The gray dots are an
        exhaustive enumeration of $\bigl(P(G), C(G)\bigr)$ for all
        $2^{K(K-1)/2}$ (equal to $64$ for $K=4$ and equal to $32\,768$ for
        $K=6$) possible ambiguity graphs $G$. Note that the trade-off curve can
        be convexified by time sharing (not shown in the figure).}

    \label{fig:tradeoff}
\end{figure}

Our design heuristic for choosing the graph $G$ is as follows. Instead of
optimizing $\bigl(P(G), C(G)\bigr)$ directly, we use the lower
bound~\eqref{eq:pcg} on $C(G)$ as a substitute. With this substitution, the
optimizing $G$ can be readily found. Observe that $P(G)$ does not decrease
(and usually increases) whenever we remove an edge from $G$. At the same
time,~\eqref{eq:pcg} stays constant whenever we remove an edge from $G$ that is
not achieving the maximum probability of pairwise error. This implies that all
optimal (for the modified criterion) ambiguity graphs have edge sets of the form
\begin{equation}
    \label{eq:egamma}
    \mc{E}_k = 
    \bigl\{ 
        k'\neq k: \Pp\bigl( (\tau_k, \omega_k) \in \mc{S}_{k,k'} \bigr) \leq
        \gamma
    \bigr\}
\end{equation}
for some threshold parameter $\gamma$ and with $\mc{S}_{k,k'}$ as defined
in~\eqref{eq:skk}. In words, all target pairs that are pairwise more difficult
to disambiguate than the threshold parameter $\gamma$ form an edge in the
ambiguity graph. Let $G_\gamma$ be the corresponding graph.  By sweeping
$\gamma$ from $0$ to $1$, we create a sequence of up to $K(K-1)/2$ optimal
graphs $G_\gamma$, which in turn trace out the detection-association trade-off
$\bigl(P(G_\gamma), C(G_\gamma)\bigr)$.  Fig.~\ref{fig:tradeoff} shows two
examples of this approach. The figure indicates that the proposed design
heuristic can be quite close to optimal.

To summarize this section, our proposed approach is as follows. Fix a value of
the threshold parameter $\gamma$ to some value between $0$ and $1$ depending on
the application requirements. Construct the corresponding ambiguity graph
$G_\gamma$ with edges given by~\eqref{eq:egamma}. Solve the corresponding
optimal ambiguity-aware beamforming problem~\eqref{eq:problem} and steer the
radar beams accordingly. From the radar returns, find all the detections and
associate them using the nearest-neighbor association rule in
Algorithm~\ref{alg:nn} parameterized by the graph $G_\gamma$.

\section{Discussion and Conclusion}
\label{sec:conclusion}

In this paper, we studied the problem of jointly designing transmit beam
patterns and association rules for tracking multiple targets using MIMO radar.
We introduced the concept of the ambiguity graph used to describe which targets
are hard to disambiguate based on prior information. We first formulated a
semidefinite program to solve the problem of designing beam patterns that avoid
simultaneous illumination of hard to disambiguate targets. We then designed the
association rule assigning detections to the different targets being tracked.
Finally, we solved the problem of designing ambiguity graphs by using a
heuristic to approximate the optimal detection-association trade-off.

In our analysis we made a number of simplifying assumptions. Some of these can
be readily relaxed. For instance, the ambiguity-aware beamforming
problem~\eqref{eq:problem} completely nulls interfering target signal
contributions, i.e., $\bm{a}^\dagger(\theta_k) \bm{R} \bm{a}(\theta_{k'}) = 0$
for all $k$ and all $k'\in\mc{E}_k$. This constraint may be too restrictive, and
it may be sufficient to ensure that the interfering target signals are below the
noise floor, i.e., $\abs{\bm{a}^\dagger(\theta_k) \bm{R} \bm{a}(\theta_{k'})}
\leq \delta$ for all $k$ and all $k'\in\mc{E}_k$ with $\delta >0$. The resulting
optimization problem remains convex and can be solved efficiently.

Similarly, we have assumed that the target azimuth angles are known a priori. To
relax this, assume that we have prior upper and lower bounds $\theta_k^-$ and
$\theta_k^+$ on each target azimuth $\theta_k$.  The ambiguity-aware beamforming
problem~\eqref{eq:problem} can be modified to accommodate this uncertainty by
replacing the objective with $\min_{k\in\{1,2,\dots,K\}}
\min_{\theta\in[\theta_k^-, \theta_k^+]} \bm{a}^\dagger(\theta) \bm{R}
\bm{a}(\theta)$ and the constraints with $\max_{\theta\in[\theta_k^-,
\theta_k^+]} \max_{\theta'\in[\theta_{k'}^-, \theta_{k'}^+]}
\abs{\bm{a}^\dagger(\theta) \bm{R} \bm{a}(\theta')} \leq \delta$ for all
$k$ and all $k'\in\mc{E}_k$. Here $\delta>0$ is a small but strictly
positive number that controls the amount of interference as discussed in the
last paragraph. This optimization problem is again convex.

There are several directions for future work. Our analysis assumed that the
ambiguity function is ideal, and we have argued how to construct corresponding
waveforms in the large-bandwidth regime. Joint beamforming and association
design under nonideal ambiguity function is an interesting open problem.
Further, while we have jointly solved the beamforming and association problems,
we have not directly considered the tracking problem.  Solving all three
problems jointly is an interesting direction for future work involving several
new aspects such as how to handle association failures and how to initiate
tracks.

\appendices

\section{Waveforms with Approximately Ideal Ambiguity Function}
\label{sec:app_corr}

We construct waveforms of support $[0,T)$ and (approximate) bandwidth $B$
that have close to ideal ambiguity function
$\bm{\chi}(\Delta\tau,\Delta\omega)$ as defined in~\eqref{eq:corr}. We
formally state the result as the following theorem.

\begin{theorem}
    \label{thm:amb}
    Let $\delta > 0$ with $1/\delta\in\N$, and let $T > 0$, $B > 0$ with $BT\in\N$, 
    satisfying
    \begin{equation}
        \label{eq:condbt}
        2^{-8}\delta^2 BT - 2\ln(BT) > \ln(2^9 N^2 /\delta^3).       
    \end{equation}
    There exists waveforms $\tilde{\bm{s}}(t)\in\C^{N}$ of support $[0,T)$ and
    approximate (i.e., Rayleigh or equivalently half the null-to-null)
    bandwidth $B$ with ambiguity function
    $\bm{\chi}(\Delta\tau,\Delta\omega)\in\C^{N\times N}$ having the
    following properties:
    \begin{enumerate}
        \setlength{\itemsep}{4pt}
        \item $\chi_{n,n}(0, 0) = 1$ \\
            for all $n\in\{1,2, \dots, N\}$.

        \item $\abs{\chi_{n,n}(\Delta\tau, \Delta\omega)} \leq \delta$ \\
            for all $n\in\{1,2, \dots, N\}$ and for all $\Delta\tau, 
            \Delta\omega\in\R$ with either $\abs{\Delta\tau} \geq 1/B$ 
            or $26/(\delta T)\leq \abs{\Delta\omega} \leq\pi  B$.

        \item $\abs{\chi_{n,n'}(\Delta\tau, \Delta\omega)} \leq \delta$ \\
            for all $n, n'\in\{1,2, \dots, N\}$ with $n\neq n'$ and for all
            $\Delta\tau, \Delta\omega\in\R$.
    \end{enumerate}
\end{theorem}

\begin{figure}
    \centering 
    \subfigure[$n = n'$]{\includegraphics{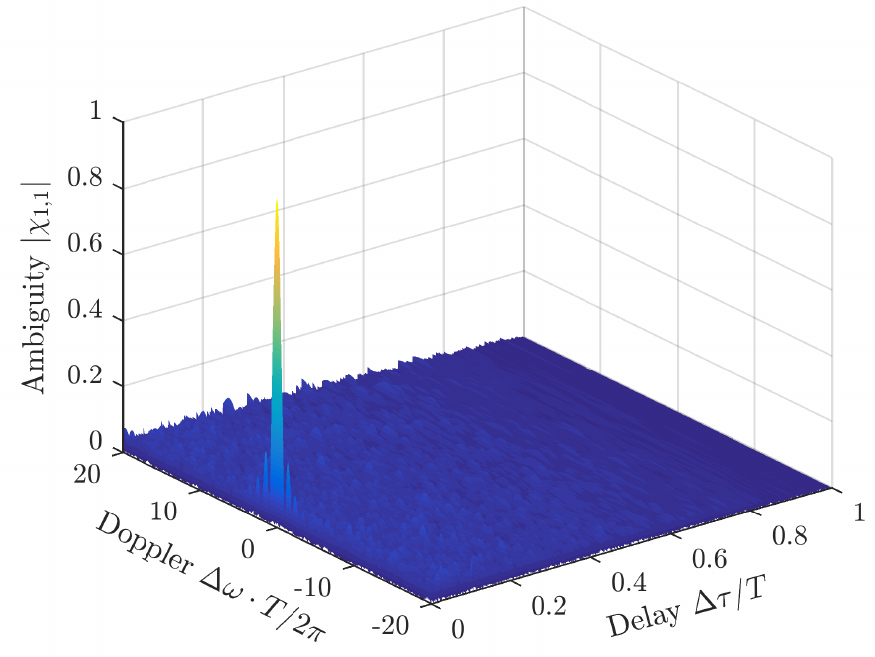}}
    \subfigure[$n \neq n'$]{\includegraphics{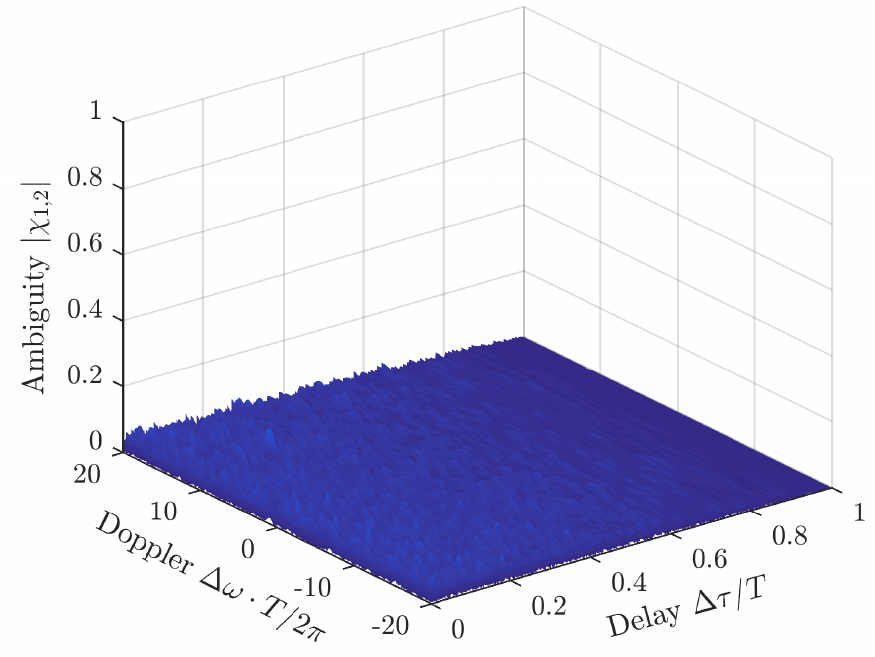}}

    \caption{Ambiguity function for randomly constructed waveforms
        $\tilde{\bm{s}}(t)$ as in Theorem~\ref{thm:amb} with temporal support $T
        = \SI{1}{\second}$ and bandwidth $B=\SI{e3}{\hertz}$. Notice that even
        for this relatively low bandwidth, the ambiguity function is already quite
        close to ideal.}
        
    \label{fig:amb}
\end{figure}

While Theorem~\ref{thm:amb} only proves existence of such a set of waveforms,
the proof shows that if $BT$ is slightly larger than the minimum necessary to
satisfy~\eqref{eq:condbt}, then the vast majority of waveforms generated by the
randomized construction described below will have the above properties. The
ambiguity function for one such set of waveforms generated at random is shown in
Fig.~\ref{fig:amb}.

\begin{example}
    \label{eg:amb}
    For a numerical example, consider temporal support $T=\SI{1}{\second}$,
    bandwidth $B=\SI{1}{\giga\hertz}$, and $N=16$ antennas. Then the
    condition~\eqref{eq:condbt} is satisfied for any side-lobe level of
    $\delta\geq\SI{-23}{\deci\bel}$. Choosing $\delta = \SI{-10}{\deci\bel}$,
    and assuming a carrier frequency of $\SI{77}{\giga\hertz}$, the limits on
    $\Delta\tau$ and $\Delta\omega$ imposed by Property~2 in
    Theorem~\ref{thm:amb} translate into a range resolution of
    $\SI{0.15}{\meter}$ and a Doppler velocity resolution of
    $\SI{0.16}{\meter/\second}$ with an upper bound of
    $\SI{1.9e6}{\meter/\second}$.
\end{example}

\begin{IEEEproof}[Proof of Theorem~\ref{thm:amb}]
As alluded to above, we use a randomized construction.
Let $c_{n,\ell}$ with $n\in\{1, 2, \dots,N\}$ and $\ell\in\{0, 1, \dots,
BT-1\}$ be independent random variables uniformly distributed over the complex
unit circle. Set
\begin{equation}
    \label{eq:polyphase}
    \tilde{s}_n(t)
    \defeq \frac{1}{\sqrt{T}}\sum_{\ell=0}^{BT-1}c_{n,\ell}\ind_{[0,1/B)}(t-\ell/B),
\end{equation}
where $\ind_{\mc{D}}(t)$ denotes the indicator function for the set
$\mc{D}\subset\R$, i.e., 
\begin{equation*}
    \ind_{\mc{D}}(t) \defeq
    \begin{cases}
        1, & \text{if $t\in\mc{D}$}, \\
        0, & \text{if $t\notin\mc{D}$}.
    \end{cases}
\end{equation*}
In words, $\tilde{s}_n(t)$ is piece-wise constant over intervals of length
$1/B$, and the value on each such interval is uniformly distributed over
the complex circle with magnitude $1/\sqrt{T}$. These waveforms therefore fall
into the class of polyphase codes~\cite[Section~4.10.2]{richards14}.
Clearly, $\tilde{s}_n(t)$ has support $t\in[0,T)$. Further, since each term
$\ind_{[0,1/B)}(t-\ell/B)$ in $\tilde{s}_n(t)$ has approximate bandwidth $B$, so
does $\tilde{s}_n(t)$ itself.

We next analyze the entries $\chi_{n,n'}(\Delta\tau,\Delta\omega)$ of the
ambiguity function.  By symmetry of the ambiguity function, we can assume
without loss of generality that $\Delta \tau \geq 0$.  The analysis proceeds in
three main steps. In the first step, we argue that all waveforms of the above
type have the desired ambiguity properties for $n'=n$ and
$\Delta\tau=\Delta\omega = 0$ and for all large enough $\abs{\Delta\tau}$ and
$\abs{\Delta\omega}$. In the second step, we argue that a set of waveforms
randomly generated as above has the desired ambiguity properties with positive
probability at all sample points $(\Delta\tau,\Delta\omega)$ of the form
$\Delta\tau = J\delta/(8B)$ and $\Delta\omega = M\delta/T $ for integer $J, M$.
Hence, there exists at least one such set of waveforms. In the third step, we
argue that if a set of waveforms has the desired ambiguity properties at all the
sample points, then it also has them at all points in between. 

Before proceeding to the main arguments we present the following two technical
lemmas that we will use throughout the proof.

\begin{lemma}
    \label{lem:chiform}
    Let $n,n'\in\{1,2,\dots,N\}$, $\varepsilon\in[0,1]$, $L\in\{0,1,2,\ldots\}$,
    and $\Delta \omega \in \R$. Then $\chi_{n,n'}$ can be expanded as
    \begin{align}
        \label{eq:chiform}
        \chi_{n,n'}\bigl((L+\varepsilon)/B, \Delta\omega\bigr)
        & = \frac{1}{T}\!\sum_{\ell=L}^{BT-1}\! c_{n,\ell}
        \bigl( 
        c_{n',\ell-L}^* a(\ell,\varepsilon,\Delta \omega)
        + c_{n',\ell-(L+1)}^* b(\ell,\varepsilon,\Delta \omega)
        \bigr),
    \end{align}
    where 
    \begin{align*}
        a(\ell,\varepsilon,\Delta\omega) 
        & \defeq \int_{t=(\ell+\varepsilon)/B}^{(\ell+1)/B} \exp(-j \Delta\omega t)\,dt, \\
        b(\ell,\varepsilon,\Delta\omega)
        & \defeq \int_{t=\ell/B}^{(\ell+\varepsilon)/B} \exp(-j \Delta\omega t)\,dt,
    \end{align*}
    and with the convention that $c_{n',-1} \defeq 0$. Furthermore,
    \begin{align*}
        \abs{a(\ell,\varepsilon,\Delta\omega)} 
        & \leq \min\bigl\{2/\abs{\Delta\omega}, 1/B\bigr\}, \\
        \abs{b(\ell,\varepsilon,\Delta\omega)} 
        & \leq \min\bigl\{2/\abs{\Delta\omega}, 1/B\bigr\}.
    \end{align*}
\end{lemma}
\begin{IEEEproof}
    We have
    \begin{align*}
        \chi_{n,n'}&\bigl((L+\varepsilon)/B, \Delta\omega\bigr) \\
        & = \int_{t=(L+\varepsilon)/B}^{T} \tilde{s}_{n}(t)
        \tilde{s}_{n'}^*\bigl(t-(L+\varepsilon)/B\bigr) \exp(-j \Delta\omega t)\,dt \\
        & = \frac{1}{T}\sum_{\ell=L}^{BT-1} c_{n,\ell}
        \biggl( 
        c_{n',\ell-L}^*\int_{t=(\ell+\varepsilon)/B}^{(\ell+1)/B} \exp(-j \Delta\omega t)\,dt 
        + c_{n',\ell-(L+1)}^*\int_{t=\ell/B}^{(\ell+\varepsilon)/B} \exp(-j \Delta\omega t)\,dt
        \biggr) \\
        & = \frac{1}{T}\!\sum_{\ell=L}^{BT-1}\! c_{n,\ell}
        \bigl( 
        c_{n',\ell-L}^* a(\ell,\varepsilon,\Delta \omega)
        + c_{n',\ell-(L+1)}^* b(\ell,\varepsilon,\Delta \omega)
        \bigr),
    \end{align*}
    proving the first part of the lemma. The second part
    follows from    
    \begin{align*}
        \biggl|\int_{t=x}^{y} \hspace{-0.3cm} \exp(-j \Delta\omega t)\,dt\biggr| 
        & = \frac{1}{\abs{\Delta\omega}} 
        \bigl|\exp(-j \Delta\omega y) - \exp(-j \Delta\omega x)\bigr| \\
        & \leq \frac{2}{\abs{\Delta\omega}} \\
        \shortintertext{and}
        \biggl|\int_{t=x}^{y} \hspace{-0.3cm} \exp(-j \Delta\omega t)\,dt\biggr| 
        & \leq \int_{t=x}^{y} \bigl|\exp(-j \Delta\omega t)\bigr|\,dt \\
        & = y-x,
    \end{align*}
    both for $x \leq y$.
\end{IEEEproof}

\begin{lemma}
    \label{lem:hoeffding}
    Let $n,n'\in\{1,2,\dots,N\}$, $\varepsilon\in[0,1]$, $L\in\{0,1,2,\ldots\}$,
    and $\Delta \omega \in \R$ as in Lemma~\ref{lem:chiform}. Then the terms in
    the expansion~\eqref{eq:chiform} of $\chi_{n,n'}\bigl((L+\varepsilon)/B,
    \Delta\omega\bigr)$ given in Lemma~\ref{lem:chiform} satisfy the following
    for any $\kappa > 0$.
    \begin{enumerate}
        \item If either $n \neq n'$ or $L \geq 1$,
            \begin{equation}
                \label{eq:aHoeffding} 
                \Pp\Biggl(
                \Biggl|\sum_{\ell=L}^{BT-1}  c_{n,\ell}
                c_{n',\ell-L}^* a(\ell,\varepsilon,\Delta \omega)\Biggr|
                > \kappa \Biggr)
                \leq 4\exp\biggl( -\frac{\kappa^2B}{4T} \biggr).    
            \end{equation}
        \item For all $n,n'$ and all $L \in \{0,1,2,\ldots\}$,
            \begin{equation}
                \label{eq:bHoeffding} 
                \Pp\Biggl(
                \Biggl|\sum_{\ell=L}^{BT-1}  c_{n,\ell}
                c_{n',\ell-(L+1)}^* b(\ell,\varepsilon,\Delta \omega)\Biggr|
                > \kappa \Biggr)
                \leq 4\exp\biggl( -\frac{\kappa^2B}{4T} \biggr).    
            \end{equation}
    \end{enumerate}
\end{lemma}
\begin{IEEEproof}
    First assume that $L \geq 1$ or $n \neq n'$. To simplify notation, define
    \begin{equation*}
        z_\ell \defeq c_{n,\ell} c_{n',\ell-L}^* a(\ell,\varepsilon,\Delta \omega).
    \end{equation*}
    By the union bound, 
    \begin{equation}
        \label{eq:reim}
        \Pp\Biggl( \Biggl|\sum_{\ell=L}^{BT-1}\!\! z_\ell\Biggr| 
        > \kappa \Biggr)
        \leq 
        \Pp\Biggl( \Biggl|\sum_{\ell=L}^{BT-1}\!\! \Re\{z_\ell\}\Biggr| 
        > \kappa/\sqrt{2} \Biggr)
        + \Pp\Biggl( \Biggl|\sum_{\ell=L}^{BT-1}\!\! \Im\{z_\ell\}\Biggr| 
        > \kappa/\sqrt{2} \Biggr).
    \end{equation}
    Since the random variables $c_{n,\ell}$ and $c_{n',\ell-L}$ are independent
    by assumption on $L$ and $n,n'$, the random variable $z_\ell$ has expected
    value zero. Moreover variables in the sequence $z_\ell$ indexed by $\ell \in
    \{L,L+1,\ldots,BT-1\}$ are mutually independent.  (This last statement can
    be observed by considering the distribution of $c_{n,\ell}c_{n',\ell-L}^*$
    conditioned on all $c_{n,\ell'}c_{n',\ell'-L}^*$ with $\ell' < \ell$, and by
    noticing that $c_{n,\ell}$ does not appear in the conditioning.)
    Furthermore, from the fact that $c_{n,\ell}$ is on the unit circle and the
    bound on the magnitude of $a(\ell,\varepsilon,\Delta \omega)$ from
    Lemma~\ref{lem:chiform}, we know that $\abs{z_\ell} \leq 1/B$. We can
    therefore apply Hoeffding's inequality~\cite[Chapter~2.6]{boucheron13} to
    both terms on the right-hand side of~\eqref{eq:reim} to obtain
    \begin{equation}
        \Pp\Biggl(
        \Biggl|\sum_{\ell=L}^{BT-1} z_\ell\Biggr\rvert
        > \kappa \Biggr)
        \leq 4\exp\biggl( -\frac{\kappa^2B}{4T} \biggr).
    \end{equation}
    This completes the proof of~\eqref{eq:aHoeffding}.

    The proof of~\eqref{eq:bHoeffding} follows similarly by noting that the
    random variable $c_{n,\ell}$ is independent of the random variables
    $c_{n',\ell-(L+1)}$ for all $n'$ and all $L\geq 0$.
\end{IEEEproof}

We start with step one of the proof, which analyzes properties of all waveforms of the
form~\eqref{eq:polyphase}. Consider first $n'=n$, $\Delta\tau = \Delta\omega =
0$. We then have
\begin{equation}
    \label{eq:step1a}
    \chi_{n,n}(0, 0)
    = \int_{t=0}^{T} \abs{\tilde{s}_n(t)}^2\,dt 
    = 1,
\end{equation}
as required. 

Consider next any $n, n'$, $\Delta\tau \geq T$ and any $\Delta\omega\in\R$. Then clearly
\begin{equation}
    \label{eq:step1b}
    \chi_{n,n'}(\Delta\tau, \Delta\omega) = 0,
\end{equation}
since each waveform $\tilde{s}_n(t)$ has support $[0,T)$.

Consider next any $n, n'$, $\Delta\tau \geq 0$, and $\abs{\Delta\omega} \geq
4B/\delta$.  Let $L \defeq \floor{\Delta\tau B}$ and $\varepsilon \defeq
\Delta\tau B - L$ so that $\Delta\tau = (L+\varepsilon)/B$. By
Lemma~\ref{lem:chiform},
\begin{align}
    \label{eq:step1dnew}
    \bigl|\chi_{n,n'} (\Delta \tau, \Delta\omega)\bigr|
    & =  \bigl|\chi_{n,n'} \bigl( (L+\varepsilon)/B, \Delta\omega \bigr) \bigr| \notag\\
    & \leq \frac{1}{T}\sum_{\ell=L}^{BT-1} 
    \bigl(\abs{a(\ell,\varepsilon,\Delta \omega)}
    + \abs{b(\ell,\varepsilon,\Delta \omega)}\bigr) \notag \\
    & \leq \frac{1}{T}\cdot BT \cdot \frac{4}{|\Delta \omega|}\notag\\
    & \leq \delta,
\end{align}
as needed. This concludes step one. 

We proceed with step two of the proof, which analyzes properties valid with
positive probability by waveforms of the form~\eqref{eq:polyphase}.  Consider
$\Delta \tau$ of the form $J \delta/(8 B)$ with $J \in
\bigl\{0,1,2,\ldots,\floor{8BT/\delta}\bigr\}$, and consider $\Delta \omega$ of
the form $M \delta/T$ with $M \in \Z$.  Let $L \defeq \floor{J \delta/8}$ and
$\varepsilon \defeq J \delta/8 - L$ so that $J \delta/(8 B) =
(L+\varepsilon)/B$. 

From Lemma~\ref{lem:chiform}, we have
\begin{align}
    \label{eq:chiineq}
    \bigl|\chi_{n,n'} \bigl(J\delta/(8 B), M\delta/T\bigr)\bigr| \notag
    & = \bigl|\chi_{n,n'}\bigl((L+\varepsilon)/B, M\delta/T\bigr)\bigr| \notag\\
    & \leq \frac{1}{T}\left|\sum_{\ell=L}^{BT-1} c_{n,\ell}
    c_{n',\ell-L}^* a(\ell,\varepsilon,M\delta/T)\right|
    + \frac{1}{T}\left|\sum_{\ell=L}^{BT-1} c_{n,\ell} c_{n',\ell-(L+1)}^* 
    b(\ell,\varepsilon,M\delta/T)\right| \notag\\
    &= \frac{1}{T}\left|\sum_{\ell=L}^{BT-1} x_\ell\right| 
    + \frac{1}{T}\left|\sum_{\ell=L}^{BT-1} y_\ell\right| 
\end{align}
with
\begin{align*}
    x_\ell & \defeq c_{n,\ell} c_{n',\ell-L}^* a(\ell,\varepsilon,M\delta/T), \\
    y_\ell & \defeq c_{n,\ell} c_{n',\ell-(L+1)}^* b(\ell,\varepsilon,M\delta/T).
\end{align*}
We now consider two cases separately, depending on the value of $J$ and $n,n'$.

\emph{Case 1: $J \geq 8/\delta$ or $n\neq n'$.} Observe that $J \geq 8/\delta$
implies $L\geq 1$. Hence, we can apply Lemma~\ref{lem:hoeffding} with $\kappa
\defeq \delta T/8$ to both terms in the right-hand side of \eqref{eq:chiineq} to
obtain
\begin{align*}
    \Pp\Biggl(
    \frac{1}{T}\Biggl|\sum_{\ell=J}^{BT-1} x_\ell\Biggr|
    > \delta/8 \Biggr)
    & \leq 4 \exp\bigl( -2^{-8}\delta^2 BT \bigr) \\
    \shortintertext{and}
    \Pp\Biggl(
    \frac{1}{T}\Biggl|\sum_{\ell=J}^{BT-1} y_\ell\Biggr|
    > \delta/8 \Biggr)
    & \leq 4 \exp\bigl( -2^{-8}\delta^2 BT \bigr).
\end{align*}
Hence
\begin{align*}
    \Pp\Biggl(
    \frac{1}{T}\Biggl\lvert\sum_{\ell=J}^{BT-1} x_\ell\Biggr\rvert 
    + \frac{1}{T}\Biggl\lvert\sum_{\ell=J}^{BT-1} y_\ell\Biggr\rvert 
    > \delta/4
    \Biggr)
    & \leq 
    \Pp\Biggl( \frac{1}{T}\Biggl\lvert\sum_{\ell=J}^{BT-1} x_\ell\Biggr\rvert > \delta/8 \Biggr)
    + \Pp\Biggl( \frac{1}{T}\Biggl\lvert\sum_{\ell=J}^{BT-1}  y_\ell\Biggr\rvert > \delta/8 \Biggr) \\
    & \leq 8 \exp\bigl( -2^{-8}\delta^2 BT \bigr).
\end{align*}
Substituting into~\eqref{eq:chiineq} yields that
\begin{equation}
    \label{eq:chiatgridL1}
    \Pp\Bigl(\bigl|\chi_{n,n'}\bigl(J \delta/(8 B), M \delta/T\bigr)\bigr|
    \leq \delta/4 \Bigr) \\
    \geq 1-8\exp\bigl( -2^{-8}\delta^2 BT \bigr).
\end{equation}

\emph{Case 2: $J < 8/\delta$ and $n=n'$.} In this case we have $L = 0$,  so that
$x_\ell = a(\ell,\varepsilon,M \delta/T)$.  Moreover, since $n =n'$, we can
restrict ourselves to $\abs{\Delta \omega} \in \bigl[26/(\delta T),\pi B\bigr]$.
We can upper bound the first term in the right-hand side of \eqref{eq:chiineq}
as
\begin{align}
    \label{eq:xlimitL1}
    \frac{1}{T}\left|\sum_{\ell=L}^{BT-1} x_\ell\right|
    &= \frac{1}{T}\left|\sum_{\ell=0}^{BT-1} a(\ell,\varepsilon,\Delta \omega)\right| \notag\\
    &= \frac{1}{T}\left|\sum_{\ell=0}^{BT-1} \int_{t=(\ell+\varepsilon)/B}^{(\ell+1)/B} \exp(-j \Delta\omega t)\,dt\right| \notag\\
    &= \frac{1}{T\abs{\Delta\omega}} \left|\sum_{\ell=0}^{BT-1} \Bigl( \exp\bigl(-j \Delta\omega \tfrac{\ell+1}{B}\bigr)- \exp\bigl(-j \Delta\omega \tfrac{\ell+\varepsilon}{B}\bigr)\Bigr)\right| \notag\\
    &=\frac{\bigl| \exp(-j\tfrac{\Delta\omega}{B})- \exp(-j\tfrac{\Delta\omega \varepsilon}{B})\bigr|}{T\abs{\Delta\omega}} \left|\sum_{\ell=0}^{BT-1}  \exp(-j \tfrac{\Delta\omega \ell}{B})\right| \notag\\
    &= \frac{\Bigl| \bigl(\exp(-j\tfrac{\Delta\omega}{B})- \exp(-j\tfrac{\Delta\omega \varepsilon}{B})\bigr) \bigl(1- \exp(-j \Delta \omega T)\bigr)\Bigr|}{T \bigl|\Delta\omega \bigl(1- \exp(-j \tfrac{\Delta \omega}{B})\bigr)\bigr|} \notag\\
    &= \frac{\bigl|(1- \exp(-j \Delta \omega T)\bigr|}{T\abs{\Delta\omega}}
    \frac{\bigl| \sin\bigl(\tfrac{\Delta\omega(1-\varepsilon)}{2B}\bigr) \bigr|}{\bigl| \sin\bigl(\tfrac{\Delta\omega}{2B}\bigr) \bigr|}.
\end{align}
Since $\abs{\Delta \omega} \leq \pi B$, we have $\Delta\omega/(2B)\in[-\pi/2,
\pi/2]$.  Using that $2 \abs{x}/\pi \leq \abs{\sin(x)} \leq \abs{x}$ for $x \in
[-\pi/2,\pi/2]$, we can thus further upper bound~\eqref{eq:xlimitL1} as
\begin{align}
    \label{eq:xlimitL0}
    \frac{1}{T}\left|\sum_{\ell=L}^{BT-1} x_\ell\right| 
    & \leq \frac{2}{T\abs{\Delta\omega}} \cdot \frac{1-\varepsilon}{2/\pi} \notag\\
    & \leq \frac{\pi}{T \abs{\Delta\omega}} \notag\\
    & \overset{(a)}{\leq} \delta\pi/26 \notag\\
    & < \delta/8,
\end{align}
where $(a)$ uses that $\abs{\Delta \omega} \geq 26/(\delta T)$ by assumption.

For the second term in the right-hand side of \eqref{eq:chiineq} we make use of
Lemma~\ref{lem:hoeffding} with $\kappa \defeq \delta T/8$ to obtain
\begin{equation}
    \label{eq:hoeffding3a}
    \Pp\Biggl(
    \frac{1}{T}\Biggl\lvert\sum_{\ell=J}^{BT-1}  y_\ell\Biggr\rvert
    > \delta/8 \Biggr)
    \leq 4 \exp\bigl( -2^{-8}\delta^2 BT \bigr).
\end{equation}

We can combine~\eqref{eq:xlimitL0} and~\eqref{eq:hoeffding3a}
with~\eqref{eq:chiineq} to yield
\begin{align}
    \label{eq:chiatgridL2}
    \Pp\Bigl( \bigl|\chi_{n,n'}\bigl(J \delta/(8 B), M \delta/T\bigr)\bigr|
    \leq \delta/4 \Bigr)
    & \geq \Pp\Biggl(
    \frac{1}{T}\Biggl\lvert\sum_{\ell=J}^{BT-1} x_\ell\Biggr\rvert 
    + \frac{1}{T}\Biggl\lvert\sum_{\ell=J}^{BT-1} y_\ell\Biggr\rvert \leq \delta/4
    \Biggr) \notag\\
    & \geq 1-4 \exp\bigl( -2^{-8}\delta^2 BT \bigr) \notag\\
    & \geq 1-8 \exp\bigl( -2^{-8}\delta^2 BT \bigr)
\end{align}
for $\abs{M \delta/T} \in [26/(\delta T),\pi B]$.

Combining~\eqref{eq:chiatgridL1} from Case~1 and~\eqref{eq:chiatgridL2} from
Case~2, we conclude that the inequality
\begin{equation}
    \label{eq:chiatgrid}
    \Pp\Bigl(\bigl|\chi_{n,n'}(\Delta\tau, \Delta\omega)\bigr|
    \leq \delta/4 \Bigr)
    \geq 1-8\exp\bigl( -2^{-8}\delta^2 BT \bigr)
\end{equation}
holds for all sample points $(\Delta \tau, \Delta \omega)$ of the form $\bigl(J
\delta/(8 B), M \delta/T\bigr)$ either when $n \neq n'$ or when $n =n'$ and
$\abs{\Delta \omega} = \abs{M \delta/T} \in [26/(\delta T),\pi B]$.

Recall from~\eqref{eq:step1b} and \eqref{eq:step1dnew} in step one 
that in this second step it suffices to consider values $\Delta\tau\in [0,T)$ and
$\Delta\omega\in(-4B/\delta, 4B/\delta)$. There are at most $8BT/\delta$ possible
values of $\Delta\tau$ of the form $J \delta/(8 B)$ in $[0, T)$, at most $8BT/\delta^2$
possible values of $\Delta\omega$ of the form $M\delta/T$ in $(-4B/\delta ,
4B/\delta)$, and $N^2$ possible values for $n, n'\in\{1,2, \dots,
N\}$.\footnote{Recall that $1/\delta$ and $BT$ are both in $\N$.}
Hence, by the union bound, with probability at least
\begin{equation*}
    1- \frac{8BT}{\delta}\cdot\frac{8BT}{\delta^2}\cdot N^2 \cdot 8\exp\bigl( -2^{-8}\delta^2 BT \bigr) \\
    = 1- \frac{2^9 B^2 T^2 N^2}{\delta^3} \exp\bigl( -2^{-8}\delta^2 BT \bigr),
\end{equation*}
the ambiguity function at those sample points $(\Delta\tau, \Delta\omega)$ has
magnitude at most $\delta/4$. If the condition
\begin{equation*}
    2^{-8}\delta^2 BT - 2\ln(BT) > \ln(2^9 N^2 /\delta^3)
\end{equation*}
holds, then this probability is strictly positive. This implies that, under this
condition, at least one set of waveforms having ambiguity function with
magnitude less than $\delta/4$ at the sample points exists, concluding step two.

For step three of the proof, take the set of waveforms with small $(\leq
\delta/4)$ ambiguity at the sample points constructed in step two. Using a
continuity argument, we will show that it has small $(\leq \delta)$ ambiguity at
all non-sample points as well.

We start with the continuity argument in the $\Delta\tau$-direction.
Consider $J\in\Z$ and assume
\begin{equation*}
    \bigl|\chi_{n,n'}\bigl(J\delta/(8B),\Delta\omega\bigr)\bigr| \leq \delta/4.
\end{equation*}
Let $\eta\in[0,1)$. We will argue that
\begin{equation*}
    \bigl|\chi_{n,n'}\bigl((J+\eta)\delta/(8B),\Delta\omega\bigr)\bigr| \leq \delta/2.
\end{equation*}
Set $L \defeq \floor{J \delta/8}$ and $\varepsilon \defeq J \delta/8 - L$, so
that $J \delta/8 = L+\varepsilon$. Since $1/\delta\in\N$, all integer
multiples of $1/B$ are also integer multiples of $\delta/(8B)$ and therefore
sample points. Since $\eta$ is strictly less than one, this implies that
$\floor{(J + \eta) \delta/8} = L$. This observation allows us to apply
Lemma~\ref{lem:chiform} to write
\begin{align*}
    \bigl|\chi_{n,n'} & \bigl(\tfrac{(J+\eta)\delta}{8B},\Delta\omega\bigr) 
    - \chi_{n,n'}\bigl(\tfrac{J\delta}{8B},\Delta\omega\bigr)\bigr| \\
    & = \bigl|\chi_{n,n'}\bigl(\tfrac{L+\varepsilon+\eta\delta/8}{B},\Delta\omega\bigr)
    - \chi_{n,n'}\bigl(\tfrac{L+\varepsilon}{B},\Delta\omega\bigr)\bigr|  \\
    & = \frac{1}{T}\Biggl|\sum_{\ell=L}^{BT-1} c_{n,\ell}
    \Bigl( 
    c_{n',\ell-L}^* \bigl(a(\ell,\varepsilon+\eta \delta/8,\Delta \omega) - a(\ell,\varepsilon,\Delta \omega)\bigr) \\
    & \hspace{1.5cm} {} + c_{n',\ell-(L+1)}^* \bigl(b(\ell,\varepsilon+\eta \delta/8,\Delta \omega) -b(\ell,\varepsilon,\Delta \omega)\bigr)
    \Bigr) \Biggr|\\
    & \leq \frac{1}{T}\sum_{\ell=L}^{BT-1} 
    \bigl|a(\ell,\varepsilon+\eta \delta/8,\Delta \omega) - a(\ell,\varepsilon,\Delta \omega)\bigr|
    + \frac{1}{T}\sum_{\ell=L}^{BT-1} 
    \bigl|b(\ell,\varepsilon+\eta \delta/8,\Delta \omega) -b(\ell,\varepsilon,\Delta \omega)\bigr| \\
    & = \frac{BT}{T}\biggl|\int_{t=\varepsilon/B}^{(\varepsilon+\eta \delta/8)/B} \exp(-j \Delta\omega t)\,dt\biggr|
    + \frac{BT}{T}\biggl|\int_{t=\varepsilon/B}^{(\varepsilon+\eta \delta/8)/B} \exp(-j \Delta\omega t)\,dt\biggr| \\ 
    & \leq B \cdot \frac{2 \eta \delta }{8 B} \\ 
    & \leq \delta/4.
\end{align*}
Hence, by the triangle inequality,
\begin{align}
    \label{eq:cont1}
    \bigl|\chi_{n,n'} \bigl((J+\eta)\delta/(8B),\Delta\omega\bigr)\bigr|
    & \leq \bigl|\chi_{n,n'}\bigl(J\delta/(8B),\Delta\omega\bigr)\bigr| + \delta/4 \notag\\
    & \leq \delta/2.
\end{align}

We continue with the continuity argument in the $\Delta\omega$-direction.
Consider $M\in\Z$ and assume
\begin{equation*}
    \abs{\chi_{n,n'}(\Delta\tau, M\delta/T)} \leq \delta/2.
\end{equation*}
Let $\eta\in[0,1)$. Then
\begin{align*}
    \bigl|\chi_{n,n'} \bigl(\Delta\tau, (M+\eta)\delta/T\bigr) 
    - \chi_{n,n'}(\Delta\tau, M\delta/T)\bigr|
    & \leq \frac{1}{T}\int_{t=0}^T \abs{1-\exp(-j\eta\delta t/T)}\,dt \\ 
    & \leq \frac{\delta}{T^2}\int_{t=0}^T t\,dt \\ 
    & = \delta/2.
\end{align*}
Hence
\begin{align}
    \label{eq:cont2}
    \bigl|\chi_{n,n'}\bigl(\Delta\tau, (M+\eta)\delta/T\bigr)\bigr|
    & \leq \abs{\chi_{n,n'}(\Delta\tau, M\delta/T)} + \delta/2 \notag\\
    & \leq \delta.
\end{align}

Together, \eqref{eq:cont1} and~\eqref{eq:cont2} show that all points
$\Delta\omega$ between sample points have small ambiguity as well. This completes
step three and the proof.
\end{IEEEproof}

\section{Number of Identifiable Targets for Complete Ambiguity Graphs}
\label{sec:app_ident}

In this appendix, we show that, for complete ambiguity graphs and for
``general'' $\bm{a}(\theta_k)$, the maximal number of identifiable targets is
$K^\star = N$.  

In Example~\ref{eg:friedlander}, we have seen an explicit construction that
allows to identify $K = N$ targets. Hence $K^\star \geq N$.

Conversely, assume that $K \geq N+1$ and consider target $k$. Since the
ambiguity graph is complete, we must have $\bm{a}^\dagger(\theta_k) \bm{R}
\bm{a}(\theta_{k'}) = 0$ for all $k'\neq k$. This implies that
$\bm{a}^\dagger(\theta_k) \bm{R} \bm{c} = 0$ for all 
$\bm{c}\in\spn\{ \bm{a}(\theta_{k'}) \}_{k'\neq k}$. We have 
$\spn\{ \bm{a}(\theta_{k'}) \}_{k'\neq k} = \C^N$ in general 
(and this is the case, in particular, for
uniform linear antenna arrays with distinct azimuths as shown in
Appendix~\ref{sec:app_ula}).  Thus, the matrix $\bm{R}$ must map the entire
space $\C^N$ into the orthogonal complement of $\bm{a}(\theta_k)$. Since this is
true for every $k$, and since $\{\bm{a}(\theta_k)\}_{k=1}^K$ again span 
$\C^N$ in general, we must have that $\bm{R} = \bm{0}$, contradicting the 
requirement $\tr(\bm{R}) = 1$.  Thus,~\eqref{eq:problem} is infeasible, 
implying that $K^\star < N+1$.

\section{Uniform Linear Antenna Arrays}
\label{sec:app_ula}

For a uniform linear antenna array with half-wavelength antenna spacing, we have
$a_n(\theta)$ as given by \eqref{eq:ula}.  Construct the matrix $\bm{A} \defeq
\bigl(\bm{a}(\theta_k)\bigr)_{k=1}^K$. Note that $\bm{A}$ is a Vandermonde matrix.
Hence, $\rank(\bm{A}) = \min\{N,K\}$ if $\{\theta_k\}_{k=1}^K$ are
distinct.

\end{document}